\documentclass[traditabstract]{aa}

\usepackage[nonamebreak]{natbib}
\usepackage[stable]{footmisc}
\usepackage{amsmath}
\usepackage{txfonts}
\usepackage{placeins}
\usepackage{natbib}
\bibpunct{(}{)}{;}{a}{}{,}
\usepackage{graphicx}
\usepackage{epstopdf}
\usepackage{ifthen}
\usepackage{mathtools}
\usepackage[breaklinks, colorlinks, citecolor=blue]{hyperref}
\usepackage{fixltx2e}
\usepackage{url}
\usepackage{hyperref}
\usepackage[table,usenames,dvipsnames]{xcolor}
\usepackage{comment} 
\usepackage{datetime} 
\usepackage{soul} 

\def\setsymbol#1#2{\expandafter\def\csname #1\endcsname{#2}}
\def\getsymbol#1{\csname #1\endcsname}

\def\Planck{\textit{Planck}}

\newbox\tablebox    \newdimen\tablewidth
\def\leaderfil{\leaders\hbox to 5pt{\hss.\hss}\hfil}

\def\endPlancktable{\tablewidth=\columnwidth 
    $$\hss\copy\tablebox\hss$$
    \vskip-\lastskip\vskip -2pt}

\def\tablenote#1 #2\par{\begingroup \parindent=0.8em
    \abovedisplayshortskip=0pt\belowdisplayshortskip=0pt
    \noindent
    $$\hss\vbox{\hsize\tablewidth \hangindent=\parindent \hangafter=1 \noindent
    \hbox to \parindent{$^#1$\hss}\strut#2\strut\par}\hss$$
    \endgroup}
\def\doubleline{\vskip 3pt\hrule \vskip 1.5pt \hrule \vskip 5pt}

\def\L2{\ifmmode L_2\else $L_2$\fi}

\def\DeltaT{\ifmmode \Delta T\else $\Delta T$\fi}
\def\deltat{\ifmmode \Delta t\else $\Delta t$\fi}
\def\fknee{\ifmmode f_{\rm knee}\else $f_{\rm knee}$\fi}
\def\Fmax{\ifmmode F_{\rm max}\else $F_{\rm max}$\fi}
\def\solar{\ifmmode{\rm M}_{\mathord\odot}\else${\rm M}_{\mathord\odot}$\fi}
\def\Msolar{\ifmmode{\rm M}_{\mathord\odot}\else${\rm M}_{\mathord\odot}$\fi}
\def\Lsolar{\ifmmode{\rm L}_{\mathord\odot}\else${\rm L}_{\mathord\odot}$\fi}
\def\inv{\ifmmode^{-1}\else$^{-1}$\fi}
\def\mo{\ifmmode^{-1}\else$^{-1}$\fi}
\def\sup#1{\ifmmode ^{\rm #1}\else $^{\rm #1}$\fi}
\def\expo#1{\ifmmode \times 10^{#1}\else $\times 10^{#1}$\fi}
\def\,{\thinspace}
\def\lsim{\mathrel{\raise .4ex\hbox{\rlap{$<$}\lower 1.2ex\hbox{$\sim$}}}}
\def\gsim{\mathrel{\raise .4ex\hbox{\rlap{$>$}\lower 1.2ex\hbox{$\sim$}}}}

\def\simprop{\mathrel{\raise .4ex\hbox{\rlap{$\propto$}\lower 1.2ex\hbox{$\sim$}}}}
\def\deg{\ifmmode^\circ\else$^\circ$\fi}
\def\pdeg{\ifmmode $\setbox0=\hbox{$^{\circ}$}\rlap{\hskip.11\wd0 .}$^{\circ}
          \else \setbox0=\hbox{$^{\circ}$}\rlap{\hskip.11\wd0 .}$^{\circ}$\fi}
\def\arcs{\ifmmode {^{\scriptstyle\prime\prime}}
          \else $^{\scriptstyle\prime\prime}$\fi}
\def\arcm{\ifmmode {^{\scriptstyle\prime}}
          \else $^{\scriptstyle\prime}$\fi}
\newdimen\sa  \newdimen\sb
\def\parcs{\sa=.07em \sb=.03em
     \ifmmode \hbox{\rlap{.}}^{\scriptstyle\prime\kern -\sb\prime}\hbox{\kern -\sa}
     \else \rlap{.}$^{\scriptstyle\prime\kern -\sb\prime}$\kern -\sa\fi}
\def\parcm{\sa=.08em \sb=.03em
     \ifmmode \hbox{\rlap{.}\kern\sa}^{\scriptstyle\prime}\hbox{\kern-\sb}
     \else \rlap{.}\kern\sa$^{\scriptstyle\prime}$\kern-\sb\fi}
\def\ra[#1 #2 #3.#4]{#1\sup{h}#2\sup{m}#3\sup{s}\llap.#4}
\def\dec[#1 #2 #3.#4]{#1\deg#2\arcm#3\arcs\llap.#4}
\def\deco[#1 #2 #3]{#1\deg#2\arcm#3\arcs}
\def\rra[#1 #2]{#1\sup{h}#2\sup{m}}

\def\dots{\relax\ifmmode \ldots\else $\ldots$\fi}
\def\WHzsr{\ifmmode $W\,Hz\mo\,sr\mo$\else W\,Hz\mo\,sr\mo\fi}
\def\mHz{\ifmmode $\,mHz$\else \,mHz\fi}
\def\GHz{\ifmmode $\,GHz$\else \,GHz\fi}
\def\mKs{\ifmmode $\,mK\,s$^{1/2}\else \,mK\,s$^{1/2}$\fi}
\def\muKs{\ifmmode \,\mu$K\,s$^{1/2}\else \,$\mu$K\,s$^{1/2}$\fi}
\def\muKRJs{\ifmmode \,\mu$K$_{\rm RJ}$\,s$^{1/2}\else \,$\mu$K$_{\rm RJ}$\,s$^{1/2}$\fi}
\def\muKHz{\ifmmode \,\mu$K\,Hz$^{-1/2}\else \,$\mu$K\,Hz$^{-1/2}$\fi}
\def\MJysr{\ifmmode \,$MJy\,sr\mo$\else \,MJy\,sr\mo\fi}
\def\MJysrmK{\ifmmode \,$MJy\,sr\mo$\,mK$_{\rm CMB}\mo\else \,MJy\,sr\mo\,mK$_{\rm CMB}\mo$\fi}
\def\microns{\ifmmode \,\mu$m$\else \,$\mu$m\fi}

\def\muK{\ifmmode \,\mu$K$\else \,$\mu$\hbox{K}\fi}
\def\microK{\ifmmode \,\mu$K$\else \,$\mu$\hbox{K}\fi}
\def\muW{\ifmmode \,\mu$W$\else \,$\mu$\hbox{W}\fi}
\def\kms{\ifmmode $\,km\,s$^{-1}\else \,km\,s$^{-1}$\fi}
\def\kmsMpc{\ifmmode $\,\kms\,Mpc\mo$\else \,\kms\,Mpc\mo\fi}
\providecommand{\sorthelp}[1]{}

\def\LCDM{$\Lambda$CDM}
\def\NHUNIT{\ifmmode {\rm \,cm^{-2}} \else $\rm \,cm^{-2}$ \fi} 

\def\wmap{\WMAP}
\def\aap{{A\&A}}

\def\apjs{{ApJ Supp.}}
\def\muKcmb{\ifmmode \,\mu$K$_{\rm CMB}$\else \,$\mu$K$_{\rm CMB}$\fi}
\newcommand{\planck}{\Planck}
\newcommand{\WMAP}{WMAP}

\newcommand{\OmegaM}{\ifmmode\Omega_{\rm M}\else $\Omega_{\rm M}$\fi}

\def\WMAP{{WMAP}}

\newcommand{\onesig}[1]{(68\%, \text{#1})}

\newcommand{\plik}{{\tt Plik}}

\newcommand{\commander}{{\tt Commander}}

\providecommand{\Planck}{\textit{Planck}}
\providecommand{\planck}{\Planck}

\providecommand{\text}[1]{\rm{#1}}

\providecommand{\muK}{\mu\rm{K}}

\providecommand{\LCDM}{{$\rm{\Lambda CDM}$}}

\newcommand{\begm}{\begin{pmatrix}}
\newcommand{\enm}{\end{pmatrix}}

\def\pmb#1{\setbox0=\hbox{#1}%
    \kern-.025em\copy0\kern-\wd0
    \kern.05em\copy0\kern-\wd0
    \kern-.025em\raise.0433em\box0}

\def\p2Y{\;_2Y}
\def\m2Y{\;_{-2}Y}
\def\beglet{
  \addtocounter{equation}{1}%
  \setcounter{parentequation}{\value{equation}}%
  \setcounter{equation}{0}%
  \def\theequation{\arabic{parentequation}\alph{equation}}%
  \ignorespaces
}
\def\endlet{
  \setcounter{equation}{\value{parentequation}}%
  \def\theequation{\arabic{equation}}%
}
\providecommand{\beglet}{\begin{subequations}}
\providecommand{\endlet}{\end{subequations}}

\newcommand{\mksym}[1]{\ifmmode {\rm #1}\else #1\fi}

\newcommand{\lowE}{\mksym{{\rm lowE}}}
\newcommand{\lowEStwo}{\mksym{{\rm lowE-S2}}}

\providecommand{\text}[1]{\rm{#1}}

\providecommand{\muK}{\mu\rm{K}}

\providecommand{\healpix}{\texttt{HEALPix}}

\providecommand{\LCDM}{{$\rm{\Lambda CDM}$}}

\newcommand\ba{\begin{eqnarray}}
\newcommand\ea{\end{eqnarray}}
\newcommand\bea{\begin{eqnarray}}
\newcommand\eea{\end{eqnarray}}

\newcommand\be{\begin{equation}}
\newcommand\ee{\end{equation}}

\newcommand{\degree}{\ensuremath{^\circ}}

\newcommand{\sroll}{{\tt SRoll}}
\newcommand{\srollone}{{\tt SRoll1}}
\newcommand{\srolltwo}{{\tt SRoll2}}

\newcommand{\simall}{{\tt SimAll}}

\newcommand{\clik}{{\tt clik}}

\newcommand{\cosmomc}{{\tt cosmomc}}
\newcommand{\camb}{{\tt camb}}
\newcommand{\noiseMC}{{\tt N+S+F-MC}}

\newcommand{\ghz}{GHz}

\begin{document}

\title{Reionization optical depth determination from\\ \Planck\ HFI data with ten percent accuracy}

\author{\small
L.~Pagano\inst{1,2,3,4}~\thanks{Corresponding author: L.~Pagano, luca.pagano@unife.it}
\and
J.-M.~Delouis\inst{5,3,6}
\and
S.~Mottet\inst{3,6}
\and
J.-L.~Puget\inst{4,2,3}
\and
L.~Vibert\inst{2,3}
}

\institute{\small
Dipartimento di Fisica e Scienze della Terra, Universit\`a degli Studi di Ferrara and INFN -- Sezione di Ferrara, Via Saragat 1, 44122 Ferrara, Italy \goodbreak
\and
Institut d'Astrophysique Spatiale, CNRS, Univ. Paris-Sud, Universit\'{e} Paris-Saclay, B\^{a}t. 121, 91405 Orsay cedex, France\goodbreak
\and
Institut d'Astrophysique de Paris, CNRS (UMR7095), 98 bis Boulevard Arago, F-75014, Paris, France\goodbreak
\and
LERMA, Sorbonne Universit\'{e}, Observatoire de Paris, Universit\'{e} PSL, \'{E}cole normale sup\'{e}rieure, CNRS, Paris, France
\and
Laboratoire d'Oc{\'e}anographie Physique et Spatiale (LOPS), Univ. Brest, CNRS, Ifremer, IRD, Brest, France\goodbreak
\and
Sorbonne Universit\'{e}, UMR7095, 98 bis Boulevard Arago, F-75014, Paris, France\goodbreak
}

\date{\vglue -1.5mm \today \vglue -5mm}

\abstract{\vglue -3mm 
We present an estimation of the reionization optical depth $\tau$ from an improved analysis of the High Frequency Instrument (HFI) data of \Planck\ satellite. By using an improved version of the HFI map-making code, we greatly reduce the residual large scale contamination affecting the data, characterized, but not fully removed, in the \Planck\ 2018 legacy release. This brings the dipole distortion systematic effect, contaminating the very low multipoles, below the noise level. On large scale polarization only data, we measure $\tau=0.0566_{-0.0062}^{+0.0053}$ at $68\%$ C.L., reducing the \Planck\ 2018 legacy release uncertainty by $\sim40\%$. Within the $\Lambda$CDM model, in combination with the \planck\ large scale temperature likelihood, and the high-$\ell$ temperature and polarization likelihood, we measure $\tau=0.059\pm0.006$ at $68\%$ C.L. which corresponds to a mid-point reionization redshift of  $z_{\rm re}=8.14\pm0.61$  at $68\%$ C.L.. This estimation of the reionization optical depth with $10\%$ accuracy is the strongest constraint to date.}


\authorrunning{Pagano et al.}

\titlerunning{Reionization optical depth determination from\\ \Planck\ HFI data with ten percent accuracy}

\maketitle
\section{Introduction}\label{sec:intro}

Cosmological recombination around redshift $z=1100$ produces a mostly neutral universe, starting the so called Dark Ages. At later stages the Universe's dark ages end with the formation of the first galaxies. The lack of Gunn-Peterson trough \citep{Gunn:1965hd,1965Natur.207..963S} in the spectra of distant quasars \citep{Rauch:1998xn,Becker:2001ee,Fan:2005es} revealed that the Universe had become almost fully reionized by redshift $z \simeq 6$ \citep{Dayal:2018hft}.

In the context of cosmological observations, Cosmic Microwave Background (CMB) generated at recombination and propagating almost freely to us, is mostly influenced by the total amount of free electrons along the line of sight, parametrized by the Thomson scattering optical depth to reionization $\tau$, one of the six parameters of the $\Lambda$CDM model. 

Reionization has mainly two effects on CMB power spectra. Firstly, it damps by a factor $e^{-2\tau}$ scalar perturbations as  generated at recombination.
 This makes the amplitude $A_s$ of the scalar perturbation highly degenerate with $\tau$ for high multipoles measurements. Secondly the re-scattering of the CMB photons on free electrons at the reionization epoch generates a bump on polarization power spectra at large angular scale. The position and height of this bump depend on the mean reionization redshift ($z_{re}$) and on the duration of the reionization transition. 
The measured quantity on the spectra at high multipoles is $A_s  e^{-2\tau}$ and thus $\delta A_s / A_s = 2 \delta \tau$. The measure of the large scale polarization allows to break the degeneracy with $A_s$ and provides directly $\tau$. A ten percent relative accuracy on $\tau$ correspond to a $1\%$ accuracy on $A_s$ if $\tau$ is about $0.05$. The direct measurement of $\tau$ on the reionzation peak is thus critical.

Although the reionization optical depth determination has been greatly improved in the last two decades, $\tau$ is still the less constrained parameter of the $\Lambda$CDM model \citep{Weiland:2018kon,planck2016-l06}. The reionization peak being visible only at very large scales ($\ell<10$) both in $EE$ and $TE$ spectra, it has been directly measured only on full sky polarized observations by space experiments. The first measurement from Wilkinson Microwave Anisotropy Probe (WMAP) \citep{kogut2003} gave $\tau=0.17 \pm 0.04$ based on $TE$ spectrum, while on the final 9yr WMAP maps \citet{hinshaw2012} reported $\tau=0.089\pm0.014$ measured on $TE$ and $EE$ spectra. \Planck\ collaboration in a re-analysis of the WMAP maps  \citep{planck2016-l05} used the \Planck\ 353GHz map as dust tracer rather then the WMAP dust template \citep{page2007}, based on the starlight-derived polarization directions and the Finkbeiner-Davis-Schlegel dust model \citep{Finkbeiner:1999aq}, lowering $\tau$ by roughly $2$ $\sigma$ to $\tau=0.062 \pm 0.012$. 

Using \Planck\ only data and Low Frequency Instrument (hereafter LFI) 70~GHz \citep{planck2016-l02} map as main cosmological channel, \Planck\ Collaboration found compatible values of $\tau=0.067\pm0.023$ in the 2015 release \citep{planck2014-a13} and $\tau=0.063\pm0.020$ in the 2018 legacy release \citep{planck2016-l05}. After the \Planck\ 2015 release \citet{Lattanzi:2016dzq} reanalyzed all the available datasets and combined LFI 2015 data with WMAP finding $\tau=0.066^{+0.012}_{-0.013}$.

All those results are obtained using the same general method, i.e. CMB maps are cleaned from foreground contamination and then the probability is directly computed on maps assuming Gaussian signal and noise \citep{Tegmark:1996,page2007,Lattanzi:2016dzq}. This relies on accurate estimation of the noise bias covariance matrix. An exhaustive review of all the measures before the \Planck\ 2018 legacy release can be found in \citet{Weiland:2018kon}. 

For the \Planck\ HFI data, being more sensitive than WMAP and LFI channels, but then more vulnerable to systematic effects, a different approach was followed by the \Planck\ Collaboration. In this case, given the difficulty of estimating reliable covariance matrices, a spectrum based likelihood was developed, acting on the cross-spectrum of 100 and 143 GHz maps. Following this approach, \citet{planck2014-a10} measured $\tau=0.055\pm0.009$ in an intermediate analysis  of HFI data after the \Planck\ 2015 release, while $\tau=0.051\pm0.009$ is reported in the \Planck\ 2018 legacy release \citep{planck2016-l05}.\footnote{A more conservative analysis based on pseudo-$\mathcal{C}_\ell$ estimator \citep{Hivon:2001jp,Tristram:2004if} is presented in \citet{planck2014-a25} which reports $\tau=0.058 \pm 0.012$.} Overall, still the major limitation is the presence of large scale systematic effects, highly reduced with respect to \Planck\ 2015 analysis but not brought below the noise level.

For a clearer global picture we report the main $\tau$ constraints to date, in the base $\Lambda$CDM model, for different large scale CMB datasets: 
\begin{eqnarray}
\tau &=& 0.089\pm0.013 \text{     Ka, Q and V with \wmap\ dust model,}\nonumber\\
\tau &=& 0.062\pm0.012 \text{     Ka, Q and V cleaned by 353 \ghz,}\nonumber\\
\tau &=& 0.063\pm0.020 \text{     LFI 70 \ghz,}\nonumber\\
\tau &=& 0.051\pm0.009 \text{     HFI 100 $\times$ 143 \ghz,}\label{eq:past_tau_values}
\end{eqnarray} 

\noindent the first value reported represents the final bound of \wmap\ Collaboration; the second is the most recent \wmap\ bound when the \planck\ 353~GHz map is used for the thermal dust cleaning; the last two values represent the \Planck\ 2018 legacy release bounds obtained using respectively LFI and HFI.

In this paper, we present an advanced approach to the \Planck\ HFI data in the attempt of reducing the systematic effects affecting the large scale polarization with the purpose of improving and solidifying the constraints of $\tau$. We upgrade the \sroll\ mapmaking algorithm introduced in \cite{planck2014-a10} (hereafter \srollone) with a new cleaning of residual distortions of the large signals, we call this new algorithm \srolltwo\ \citep{Delouis:2019bub}.

The paper is organized as follow: in Section~\ref{sec:mapmaking}, we present the improved mapmaking algorithm. In Sections~\ref{sec:power_spectrum} and \ref{sec:likelihood}, we present the power spectra, the main result on $\tau$, and the consistency tests performed. Finally, in Section~\ref{sec:cosmology}, we show the impact of the new $\tau$ constrain on the cosmological scenario.

\section{\label{sec:mapmaking}Map-making improvements}

The 2018 legacy HFI maps \citep{planck2016-l03} represent a great  step forward in the attempt of cleaning systematic effects contaminating the large scale polarization. In particular, the  impact of the non-linearities of the analogue-to-digital converters (ADCs) of readout chains has been substantially reduced, introducing variation in the gain of bolometer readout chains. This correction accounts for the first-order approximation of the ADC non-linearity (ADCNL) systematic effect, but still, large signals, such as foregrounds on the Galactic plane and dipoles, are affected by the second-order ADCNL effect. This was not treated by the \Planck\ Collaboration (see Section 5.13 of \citet{planck2016-l03} for details), leaving large scale residuals in polarization mainly due to a mismatch that, violating the stationarity of the signal in a given pixel, causes temperature to polarization dipole leakage.

For the analysis presented in this paper, we improve the \srollone\ code in what we call \srolltwo, in order to further reduce the polarization leakage due to strong signals. In the following, we list the main modifications introduced in the \srolltwo\ code, for more details, see \citet{Delouis:2019bub}:

\begin{itemize}
\item New ADCNL correction is obtained by fitting the residuals with a bi-dimensional spline model per bolometer as a function of signal value and time.
This  solution  removes  the  apparent  gain variation of bolometers allowing to fit only one gain for the entire mission. As  demonstrated in \citet{Delouis:2019bub}, time variation is only necessary to capture the ADCNL at 143\,GHz, and thus for the 100 and 353\,GHz bolometers, only a mono-dimensional spline is considered. We verify that, for those channels, opening the time variation does not improve substantially  the solution.
\item Internal fit of (and subsequently marginalization over) the polarization angle and polarization efficiency per bolometer.
\item Update of the CO template  based on 2015 \Planck\ release, used for the bandpass mismatch fit, introducing two new templates based on $^{12}$CO and $^{13}$CO  extracted as described in \citet{planck2016-l03} section 3.1.3 and in \citet{Delouis:2019bub} section 4.1.
\item Update of the thermal dust template  using  a map based on 2018 legacy release (for details, see section 4.1 of \citet{Delouis:2019bub}).
\item Update of the real part of the empirical transfer function used at 353\,GHz, replacing the $3$ real harmonic ranges of the spin frequency used in the \Planck\ 2018 legacy release (see \citet{planck2016-l03} section 2.2.2) by a single $10$s time constant (for details see section 4.2.2 of \citet{Delouis:2019bub}).
\end{itemize}

\begin{figure}[h]
\includegraphics[width=0.24\textwidth]{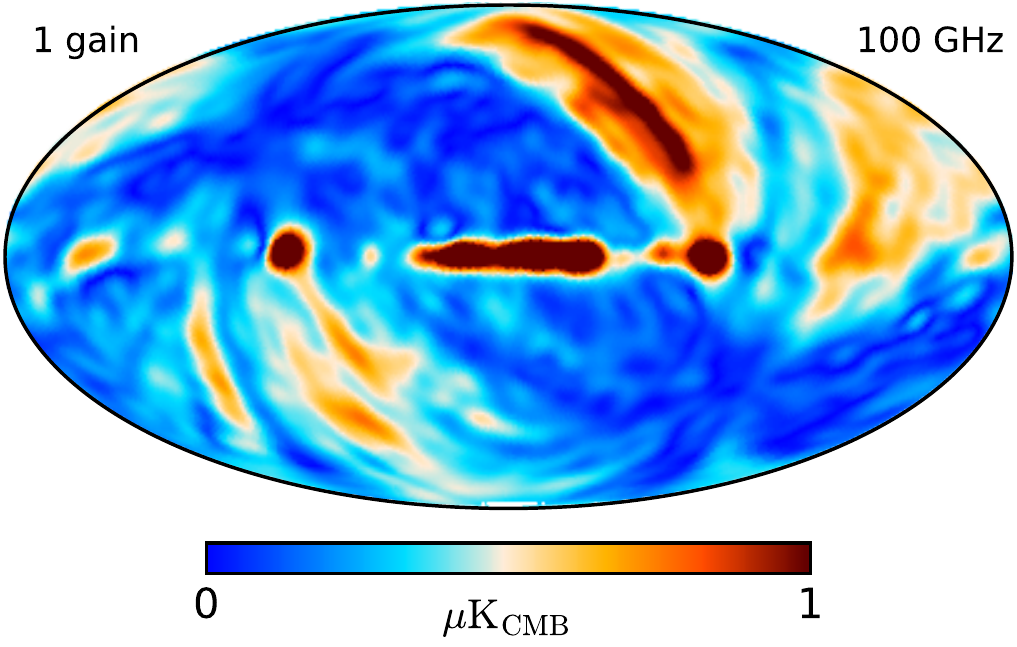}
\includegraphics[width=0.24\textwidth]{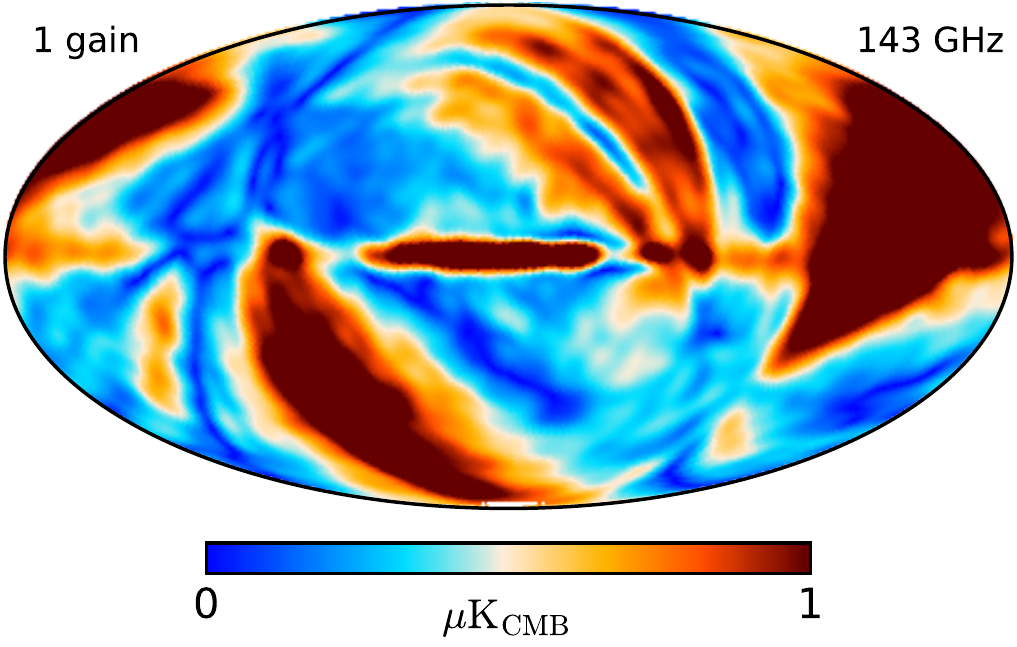}\\
\includegraphics[width=0.24\textwidth]{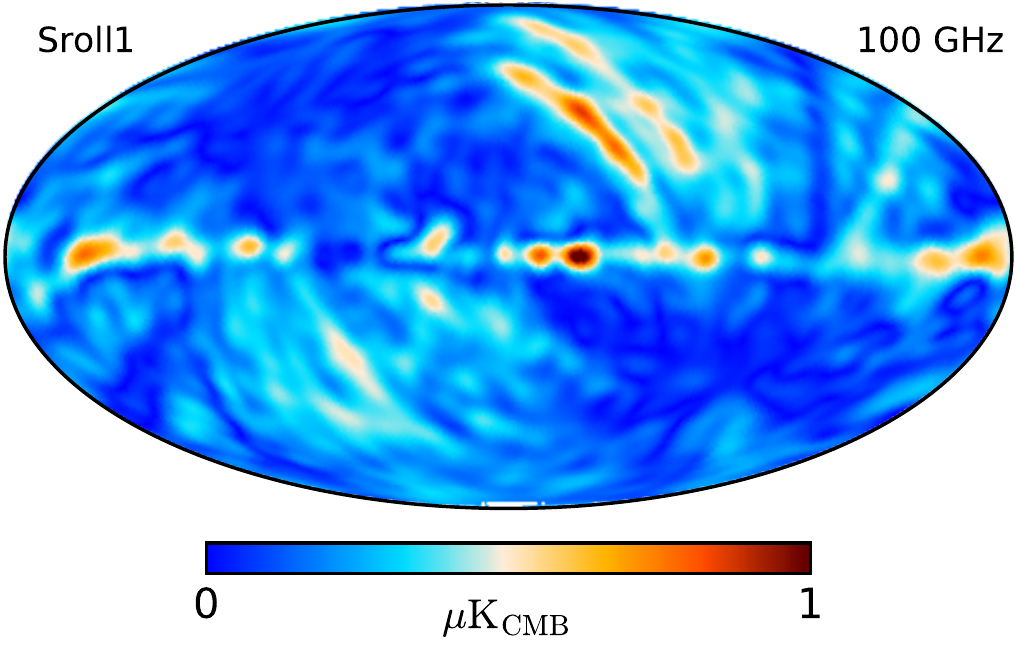}
\includegraphics[width=0.24\textwidth]{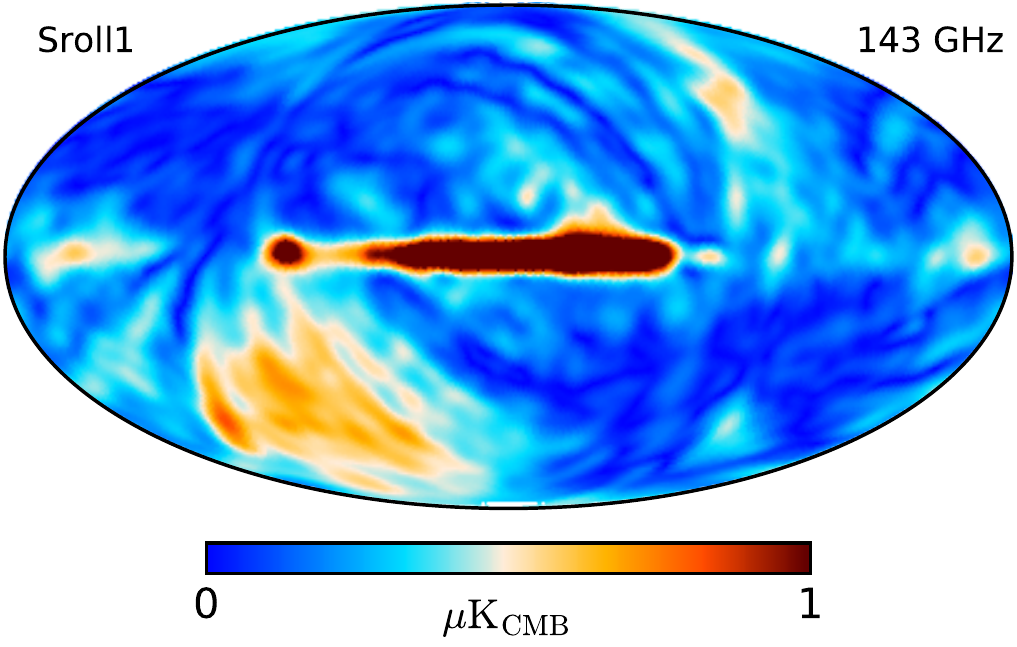}\\
\includegraphics[width=0.24\textwidth]{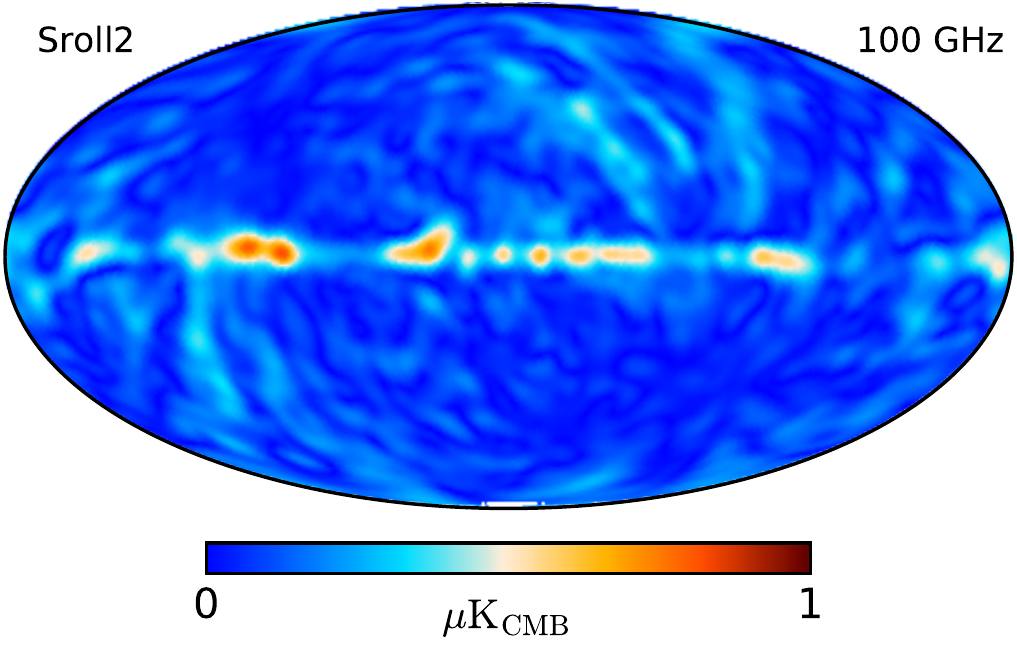}
\includegraphics[width=0.24\textwidth]{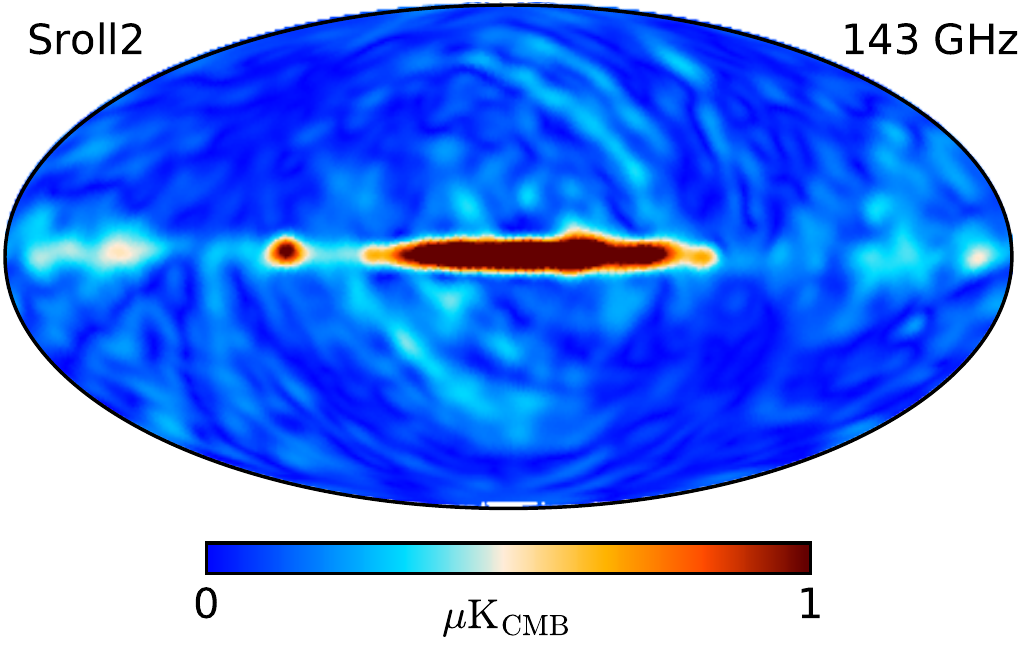}
\caption{Polarization intensity maps at 100 and 143 GHz obtained applying \srollone\ and \srolltwo\ to a set of simulated timelines. The input sky has been subtracted after the map projection. The simulated timelines contain dipole, sky signal, systematic effects and electronic noise only. The first row shows maps obtained running \srollone\  with only one gain for the entire mission, the middle row shows \srollone\ with 128 gain steps, as used in \Planck\ 2018 legacy release, and last one shows \srolltwo\ results.
\label{fig:sims}}
\end{figure}

Figure~\ref{fig:sims} shows polarization intensity maps (defined as $P\equiv\sqrt{Q^2+U^2}$ ) at 100 and 143\,GHz obtained simulating realistic sky signal affected by ADCNL and projected with \srollone\ and \srolltwo\ codes. From those maps, we remove the input sky. In the first row, we show maps obtained with \srollone\ without gain variation. In the middle row, the maps are obtained using \srollone\ opening the gain variation and fitting 128 gain steps as done in \citet{planck2016-l03}. In the last row, the simulated timelines are projected into maps with the \srolltwo\ code. The large scale dipole leakage present in the first panel is substantially reduced by the introduction of gain variability (second panel) which still leaves $\sim 1\mu {\rm K}$ level residuals as already shown \citet{planck2016-l03}. This residual is further reduced by \srolltwo\ demonstrating that the ADC-NL correction allows to fit a single gain for the bolometers as expected from pre-flight analysis \citep{holmes2008,pajot2010,catalano2010}.

 \begin{figure}[h]
 \includegraphics[width=0.48\textwidth]{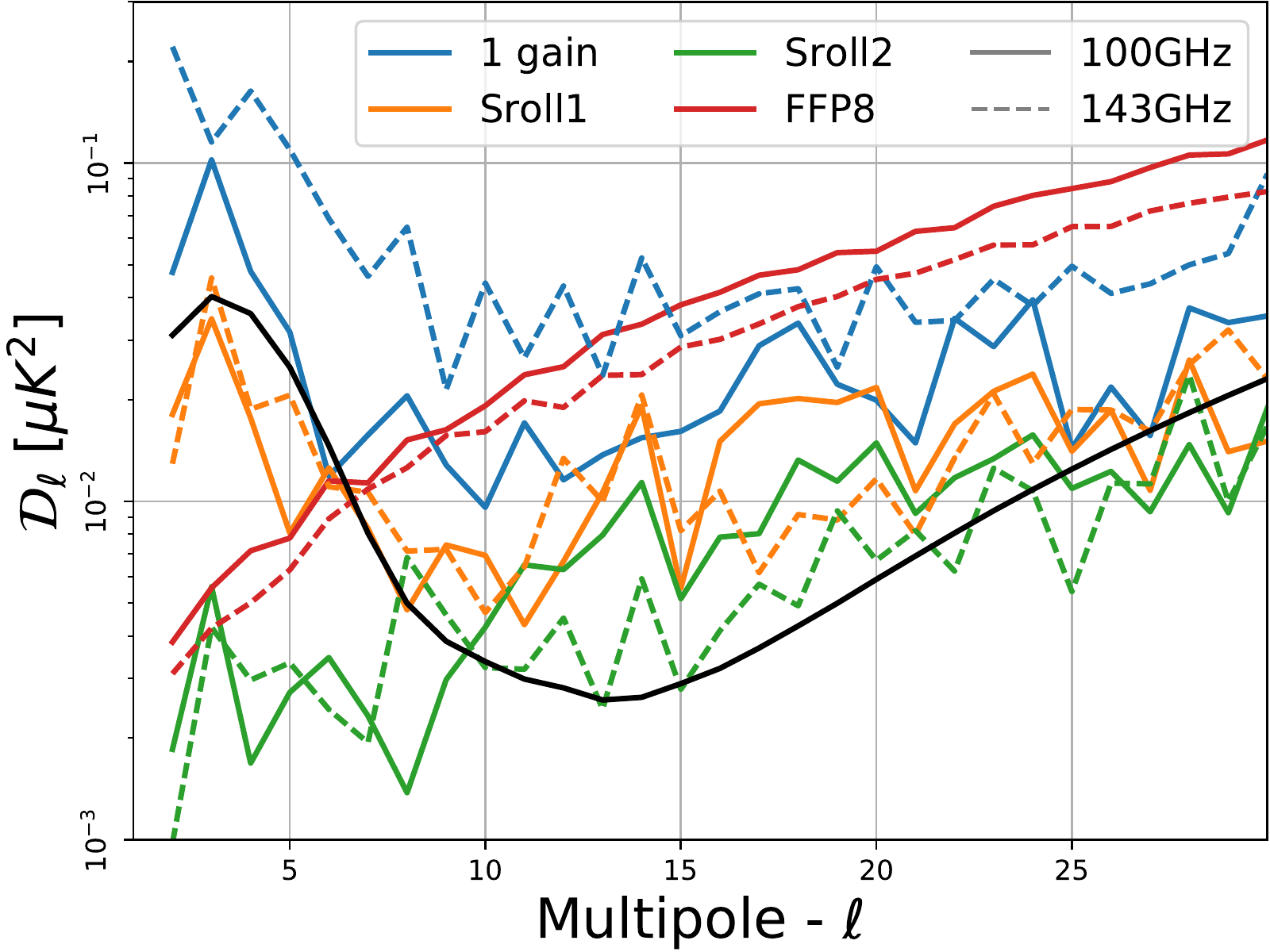}
 \caption{$EE$ pseudo auto-spectra evaluated for 100\,GHz (solid) and 143\,GHz (dashed) on the simulations shown in Fig.~\ref{fig:sims}. Here we have applied a symmetric Galactic cut of 20\,\deg, retaining 66\% of the sky. The rise at higher multipoles is caused by the autocorrelation of the electronic noise present in the maps which is not debiased. The red lines represent the average of 100 FFP8  simulations \citep{planck2014-a14} containing only white noise and $1/f$ noise. The black solid line corresponds to a $EE$ power spectrum with $\tau=0.055$.\label{fig:spectrum_sims}} 
 \end{figure}

\begin{figure}[h]
 \includegraphics[width=0.48\textwidth]{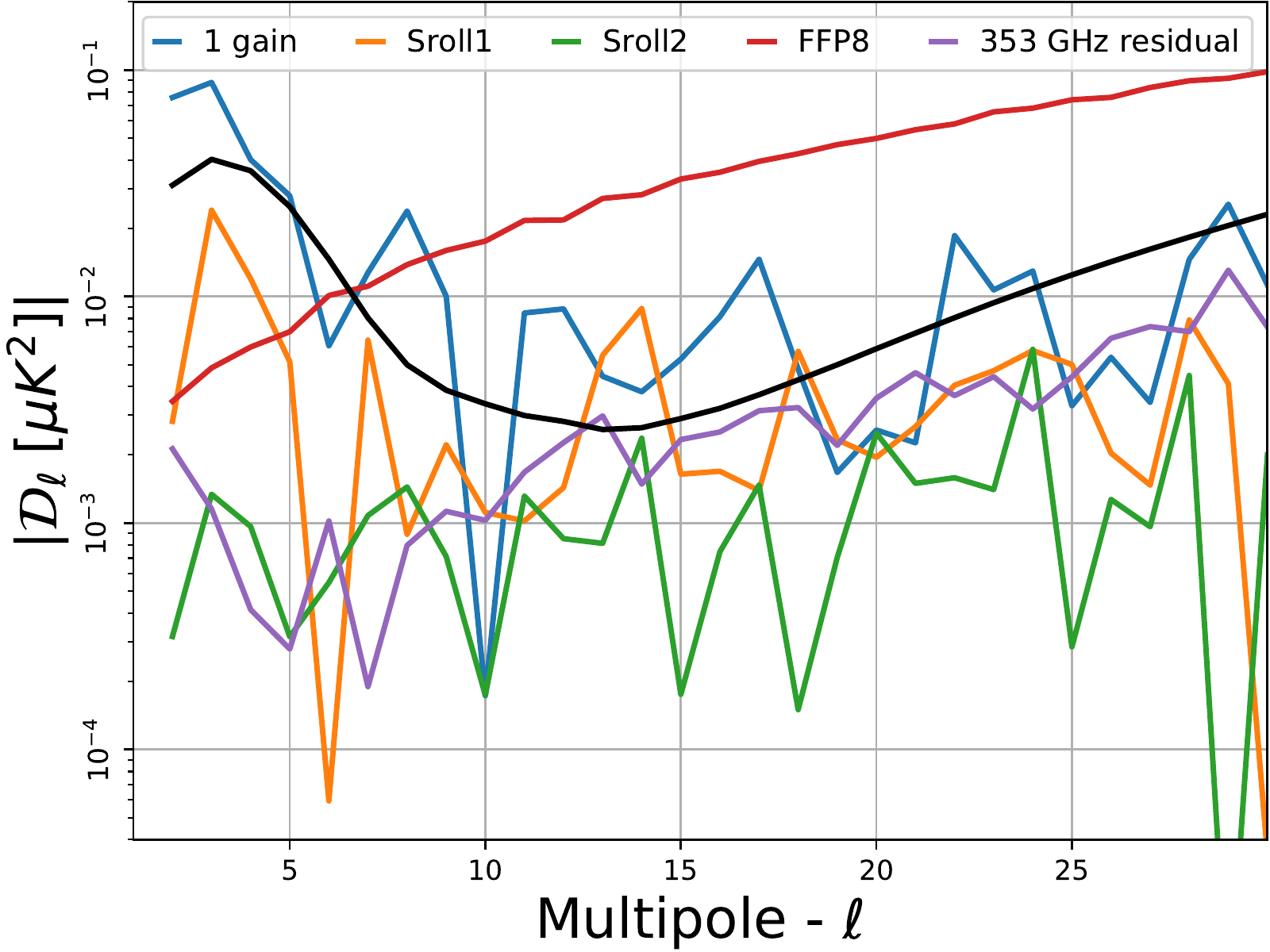}
 \caption{As Fig.~\ref{fig:spectrum_sims} but for the pseudo $100\times143$ cross-spectra. In purple, we plot the auto-spectrum of 353\,GHz residual systematic effects rescaled to $100\times143$ ($\sim 8\times 10^{-4}$ factor applied, $\sim 0.02$ from 100\,GHz and $\sim 0.04$ from 143\,GHz). The red line is the square root of the product of 100 and 143\,GHz noise spectra that is proportional to the variance associated with the noise in the cross-spectrum. In the \srolltwo\ maps, the large scale is dominated by signal and $1/f$ noise rather than residual systematic effects.\label{fig:cross_spectrum_sims}}
 \end{figure}

In Figure~\ref{fig:spectrum_sims}, we report the $EE$ pseudo power spectra ($\mathcal{D}_\ell \equiv \ell(\ell+1)\mathcal{C}_\ell/2\pi$) of the residual systematic effects (hereafter systematics) maps shown in Fig.~\ref{fig:sims}. The level of those residuals is pushed below $2\times 10^{-2} \mu {\rm K}^2$ both for 100 and 143\,GHz by \srolltwo, gaining an order of magnitude with respect to \srollone\ results. Furthermore, those residuals are weakly correlated, as can be seen in Fig.~\ref{fig:cross_spectrum_sims}. In the $100\times143$ $EE$ cross-spectrum, the level of systematics is  further reduced  below $3\times 10^{-3} \mu {\rm K}^2$ (green line of Fig.~\ref{fig:cross_spectrum_sims}). With \srolltwo, systematics are negligible with respect to a typical CMB power spectrum and below the gaussian noise level
\footnote{The noise spectrum shown in Fig.~\ref{fig:cross_spectrum_sims} should not be interpreted as noise level biasing the cross-spectrum estimate, by definition unbiassed, but as noise contribution entering, together with the theoretical CMB spectrum, in the error computation.}.

 Besides, we start to be limited by the quality of the dust template, given that the level of residual systematic effects in the $100\times143$ spectrum (green line of Fig.~\ref{fig:cross_spectrum_sims}) is below or at most equal to the systematics still present in the 353 GHz channel used as dust template for both 100 and 143 GHz (purple line).

\begin{figure}[h]
\includegraphics[width=0.24\textwidth]{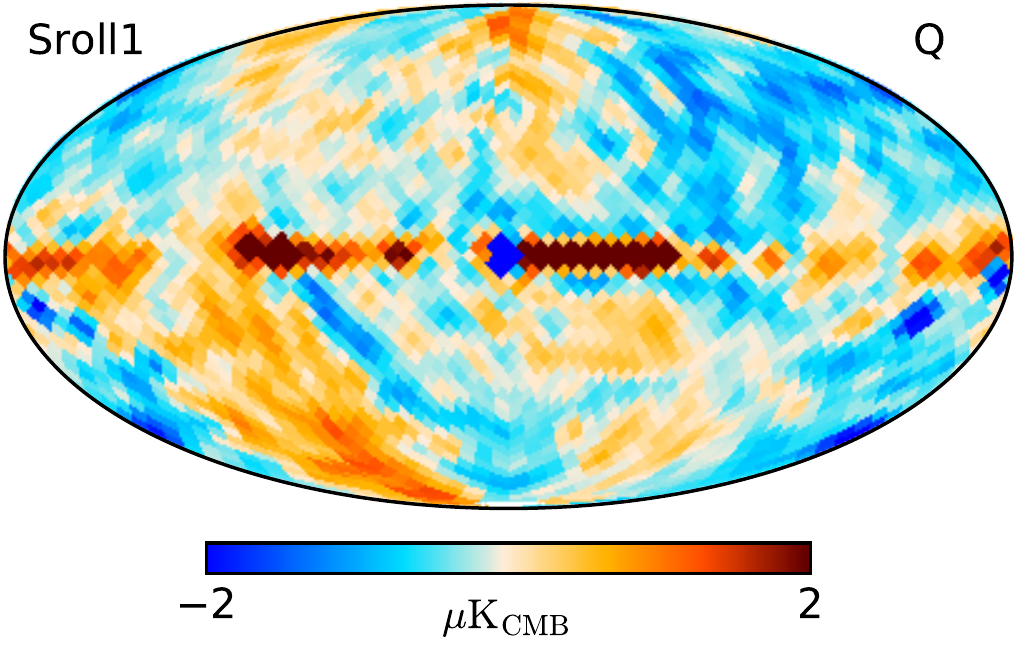}
\includegraphics[width=0.24\textwidth]{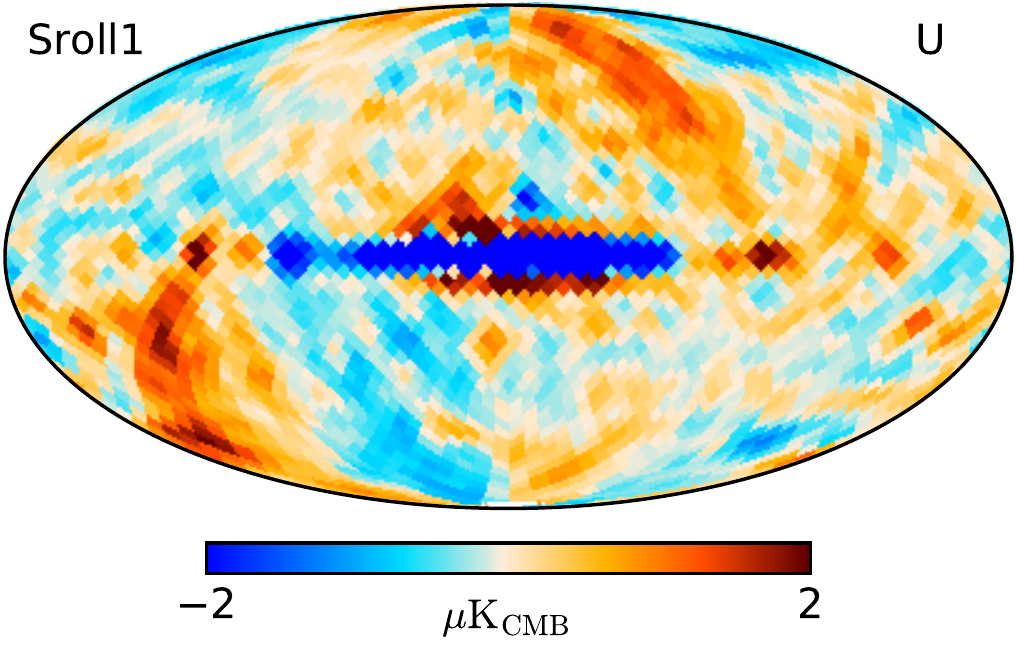}\\
\includegraphics[width=0.24\textwidth]{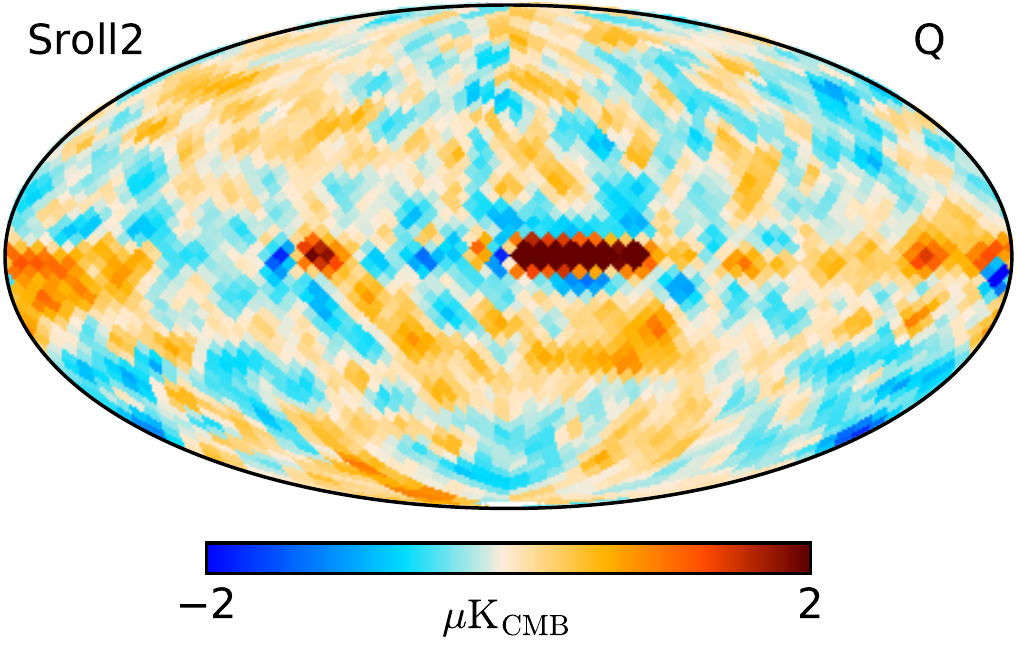}
\includegraphics[width=0.24\textwidth]{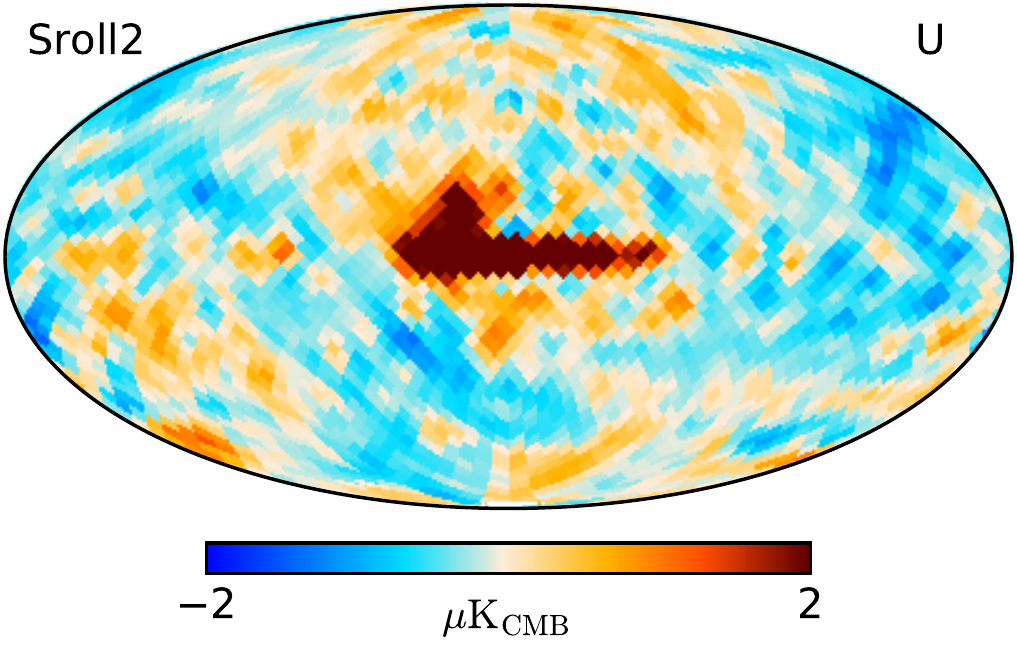}
\caption{Data Q and U maps at 100\,GHz cleaned from synchrotron and dust emissions. Top row shows the \Planck\ 2018 legacy release computation obtained with \srollone, while the bottom one the \srolltwo\ computation. \label{fig:data100}}
\end{figure}

\begin{figure}[h]
\includegraphics[width=0.24\textwidth]{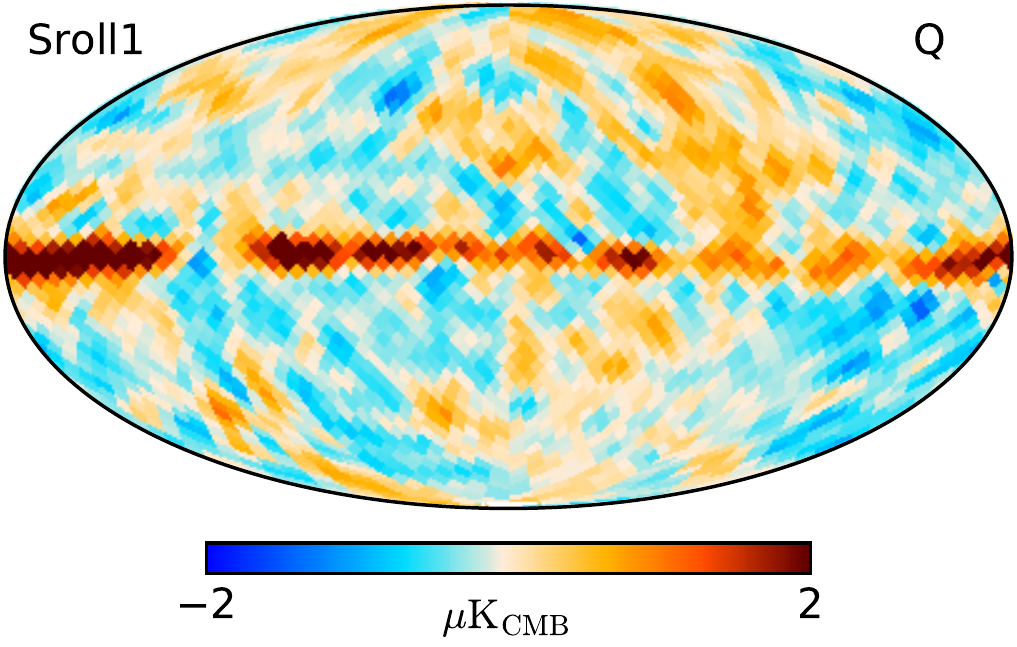}
\includegraphics[width=0.24\textwidth]{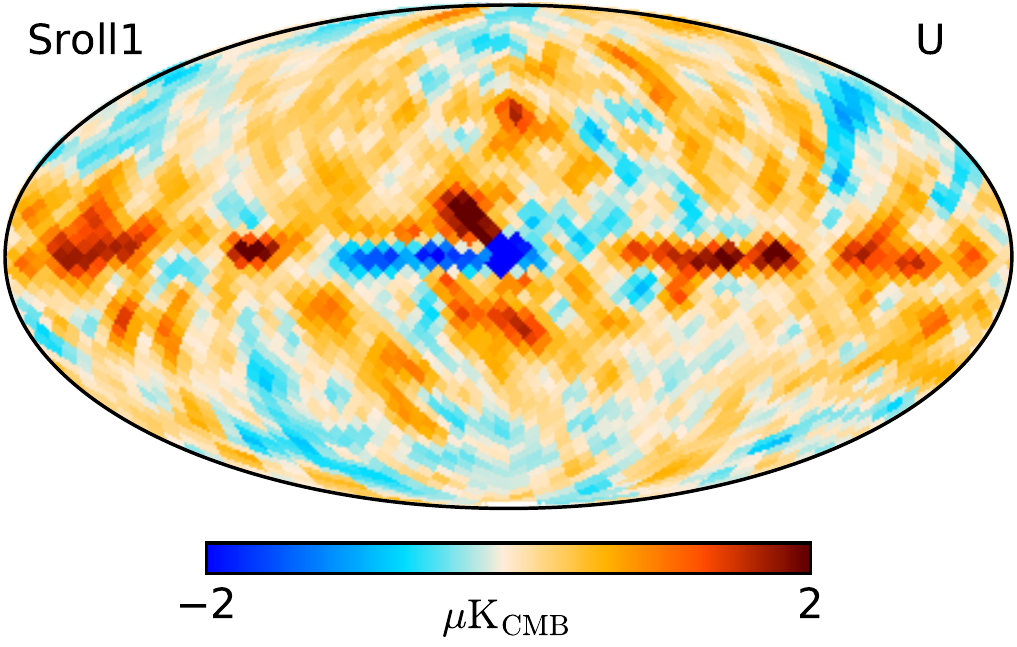}\\
\includegraphics[width=0.24\textwidth]{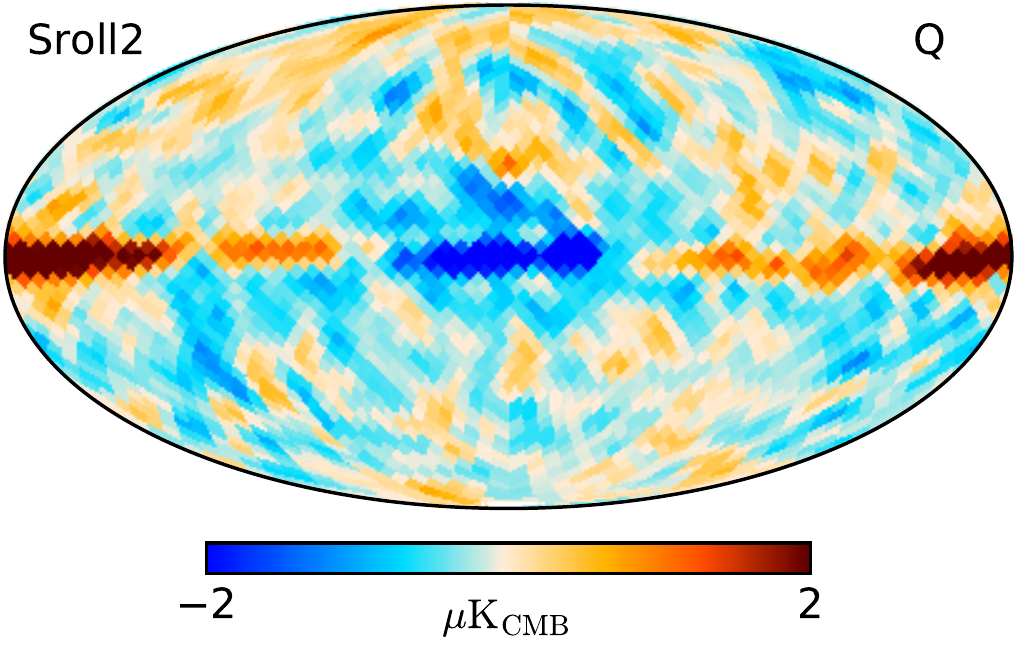}
\includegraphics[width=0.24\textwidth]{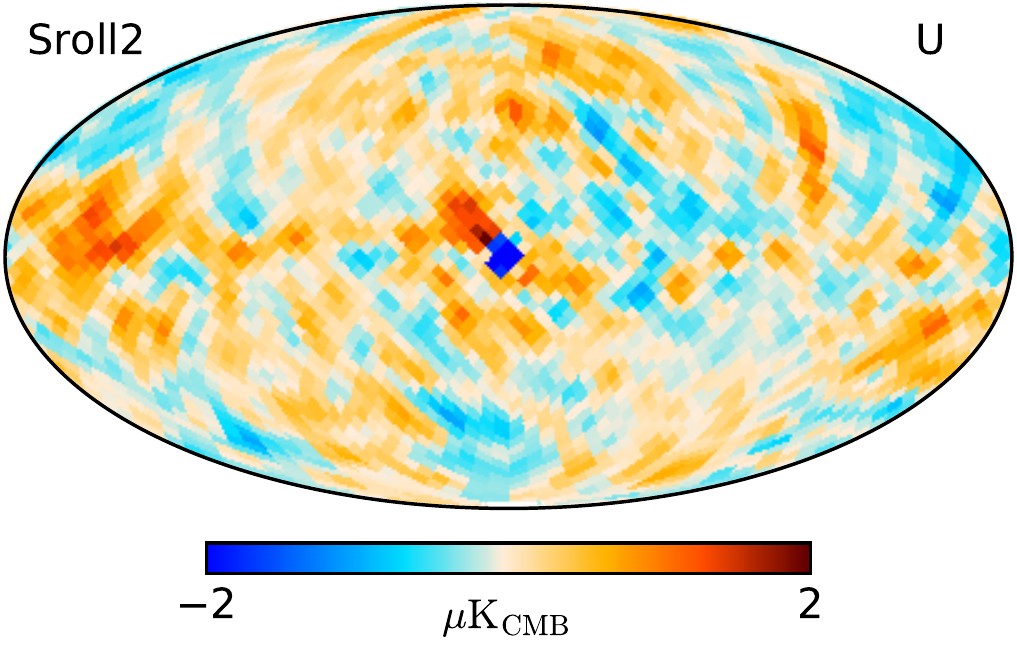}
\caption{Data Q and U maps at 143 GHz cleaned from synchrotron and dust emissions. Top panel shows the \Planck\ 2018 legacy release computation obtained with \srollone, while the bottom one the \srolltwo\ computation. \label{fig:data143}}
\end{figure}

Similar improvement is easily recognizable in \srolltwo\ maps of data. In Figures~\ref{fig:data100} and \ref{fig:data143}, we show 100 and 143\,GHz maps after the removal of diffuse Galactic foreground contamination for both \srollone\ and \srolltwo. The overall level of systematics is significantly reduced everywhere in the maps by \srolltwo. Both the large scale spurious structures associated to dipole leakage and the Galactic disc residuals are substantially improved.

\section{Power spectrum}\label{sec:power_spectrum}

\begin{figure}[h!]
 \includegraphics[width=0.48\textwidth]{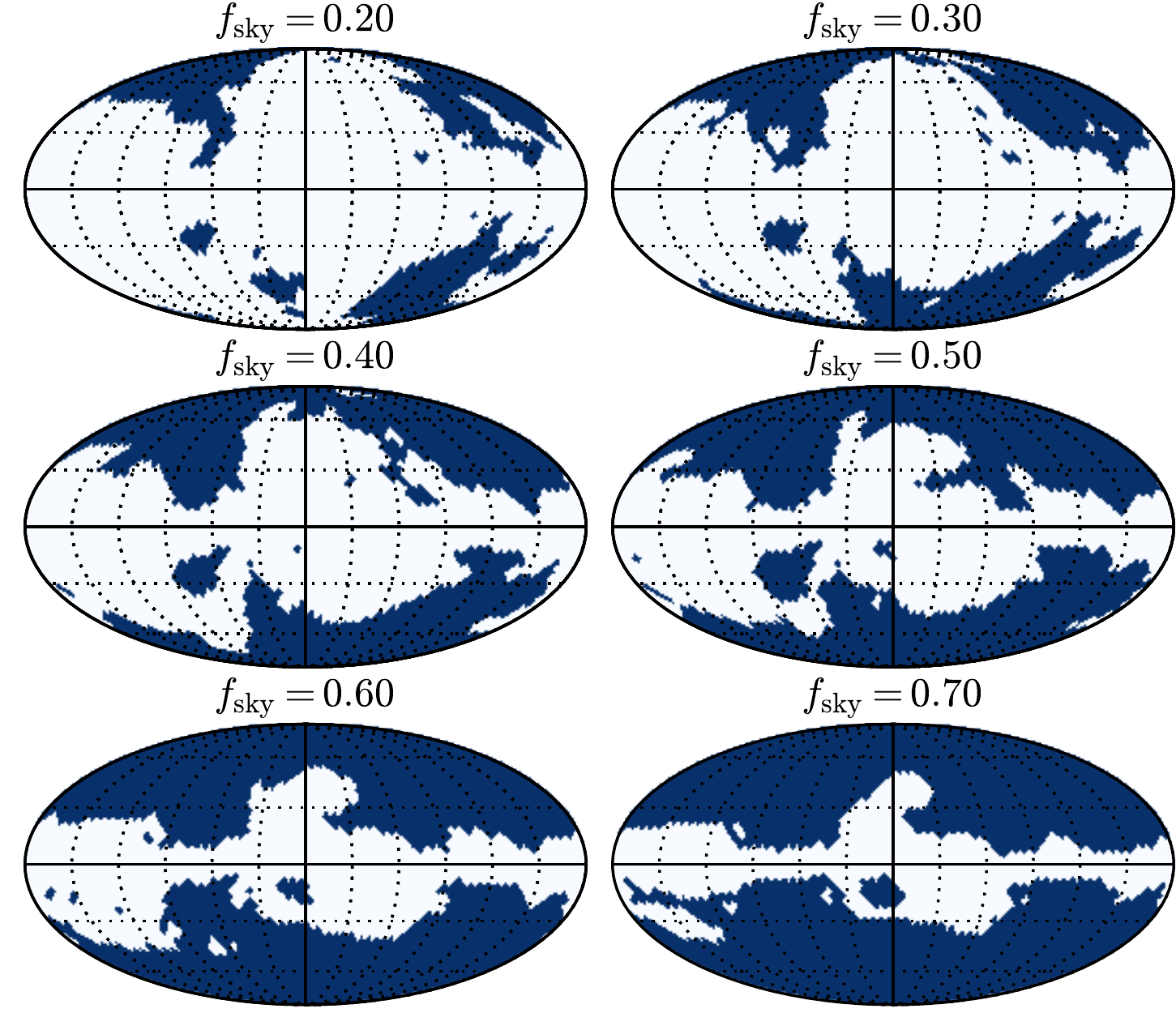}
\caption{Masks used for the present analysis. The 70\% mask is used for the foreground cleaning, the others in the cosmological analysis. All the masks used in this analysis are binary maps, without any apodization applied.
\label{fig:masks}}
 \end{figure}

\begin{figure}[h!]
 \includegraphics[width=0.48\textwidth]{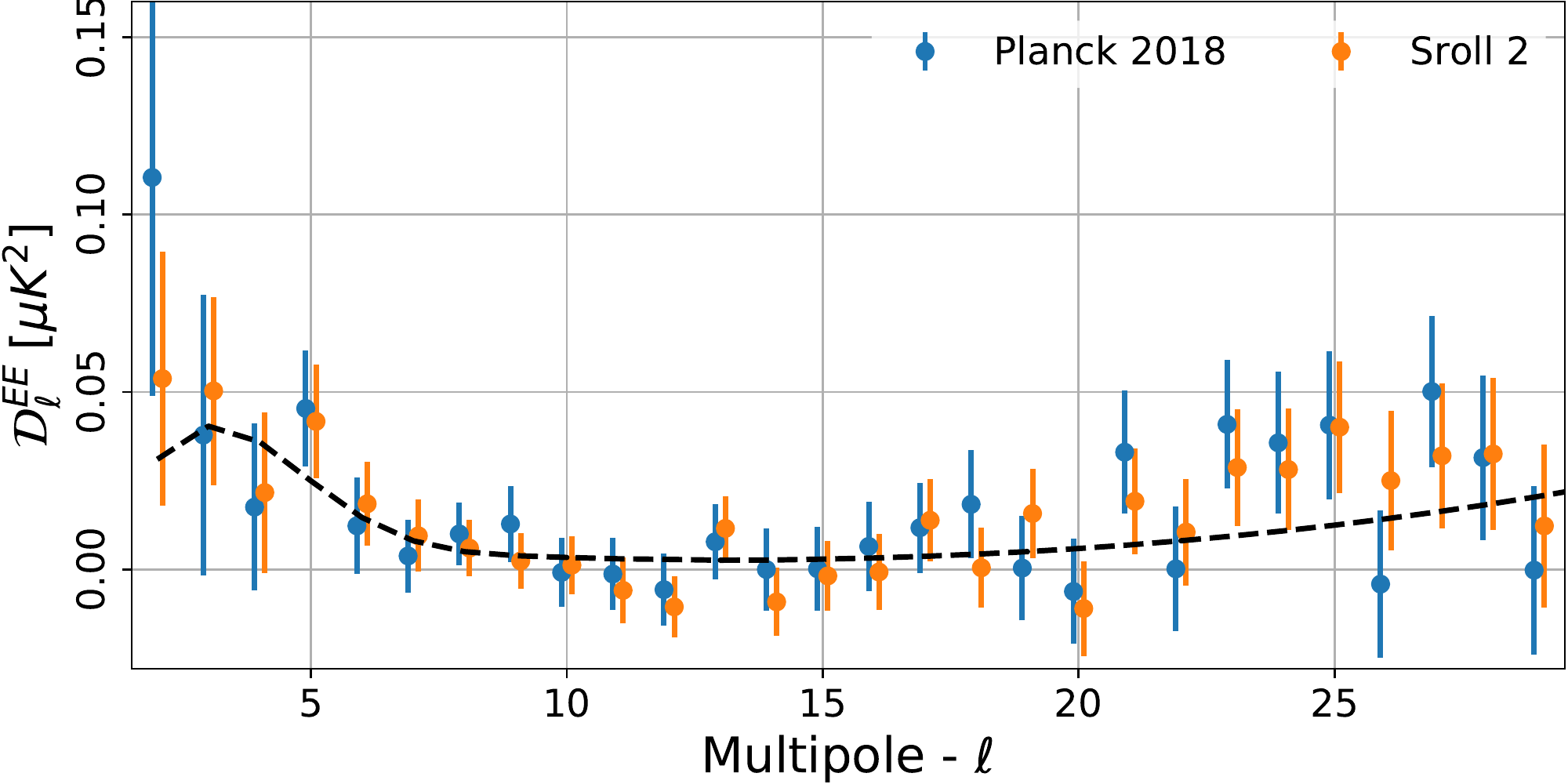}
 \caption{Low-$\ell$ $EE$ cross spectrum 100x143 for the \Planck\ 2018 legacy release (blue points) and for the \srolltwo\ maps (orange). The mask used retains 50\% of the sky. The error bars are Monte Carlo based and do include cosmic variance. The black line corresponds to a $EE$ power spectrum with $\tau=0.055$. 
 \label{fig:spectrum_comparison_2018}}
 \end{figure}

This section describes the power spectrum computation made using \srolltwo\ maps and the analysis performed on simulations. As a general approach, we follow the procedure adopted for HFI low-$\ell$ analysis presented in \citet{planck2016-l05} (section 2.2). In short, 100 and 143 GHz maps, built only using polarization sensitive bolomenters (PSBs), undergo the following procedure:
\begin{itemize}
\item We filter them with a cosine window function \citep{Benabed:2009af,planck2014-a10}, downgrading to \healpix\ \citep{gorski2005} $\rm N_{side}=16$ resolution. In order to maintain the covariances invertible, we add $20\,n{\rm K}$ of diagonal regularization noise. 
\item We remove the Galactic foreground contamination through template fitting. We employ \srolltwo\ 353 GHz map for thermal dust removal and, \WMAP\ 9yr K and Ka bands  for synchrotron at 100 and 143\,GHz respectively. The scalings found are reported in Tab.~\ref{tab:scalings}
\item We compute the cross-QML \citep{planck2016-l05,Tegmark:1996,Tegmark:2001zv,Efstathiou:2006} power spectrum between 100 and 143\,GHz cleaned maps (see Fig.~\ref{fig:spectrum_comparison_2018}) outside a Galactic mask (see Fig.~\ref{fig:masks}). As temperature map, we use the \Planck\ 2018  \commander\ solution \citep{planck2016-l04,planck2016-l05} smoothed with a 440\,\arcmin ($\sim$7.3\,\degree) gaussian beam and regularized with $2\,\mu{\rm K}$ diagonal noise.  As covariance matrices, we use FFP8 covariances \citep{planck2014-a09,planck2014-a14} for the HFI channels, and for WMAP K and Ka, the official 9yr matrices \citep{bennett2012}.
\end{itemize}

\begin{table}[htbp!]
\begingroup
\caption{Template scalings measured on data. The synchrotron tracers are \wmap\ K and Ka bands for 100 and 143\,GHz respectively. The dust tracer is the 353\,GHz map.}
\label{tab:scalings}
\nointerlineskip
\vskip -3mm
\setbox\tablebox=\vbox{
   \newdimen\digitwidth
   \setbox0=\hbox{\rm 0}
   \digitwidth=\wd0
   \catcode`*=\active
   \def*{\kern\digitwidth}
   \newdimen\signwidth
   \setbox0=\hbox{+}
   \signwidth=\wd0
   \catcode`!=\active
   \def!{\kern\signwidth}
\halign{\hbox to 1.0in{#\leaderfil}\tabskip=1em&
  \hfil#\hfil\tabskip=2em&
  \hfil#\hfil\tabskip=0pt\cr
\noalign{\doubleline}
\noalign{\vskip -2pt}
\omit\hfil Channel [GHz]\hfil& $\alpha\times10^{2}$& $\beta\times10^{2}$\cr
\noalign{\vskip 3pt\hrule\vskip 5pt}
100& $0.95\pm0.07$& $1.86\pm0.015$\cr
143& $1.63\pm0.21$& $0.0394\pm0.014$\cr
\noalign{\vskip 5pt\hrule\vskip 3pt}}}
\endPlancktable
\endgroup
\end{table}

The same cleaning procedure is applied to a set of 500 Monte Carlo simulations containing realistic sky signal, noise and systematic effects. The fiducial CMB map contained in those simulations is removed after the foreground cleaning leaving only maps with noise, systematic effects and foreground residuals, referred hereafter as \noiseMC\ (for Noise, Systematics and Foreground residuals Montecarlo) . 

As already stated in \citet{planck2014-a10} and \citet{planck2016-l05}, FFP8 covariance matrices \citep{planck2014-a14} represent a sub-optimal, but unavoidable, choice. The FFP8 covariance matrices are built following the method presented in \citet{keskitalo2010} (see in particular section 3.2). They are assembled in two pieces, one describing the sub-baseline correlation part which is untouched by the destriper mapmaking and one describing the ring-to-ring correlation due to errors in solving for the baselines. Consequently those matrices do not capture properly the variance of the systematic effects but only the white and $1/f$ noise components, assuming an analytical model for the noise spectrum\footnote{FFP8 covariance matrices can be obtained upon specific request to the Planck Project Scientist at ESTEC or directly to NERSC.}.
Despite that, since we rely on cross-spectrum estimator, this choice does not impact power spectrum estimate but only its optimality (see e.g. \citet{planck2014-a10,planck2016-l05}). 

Furthermore, for \srolltwo\ maps, having the residual systematics further reduced with respect to noise (see Fig.~\ref{fig:cross_spectrum_sims}), we have a more efficient estimator than the analysis performed for \Planck\ 2018 legacy release.

All masks used for foreground cleaning (see Fig.~\ref{fig:masks}), power spectrum estimation and likelihood are obtained thresholding the sum of dust polarization intensity scaled at 143\,GHz with synchrotron polarization intensity scaled at 100\,GHz, both smoothed with a Gaussian window function with full with half maximum of  7.5\,\deg . As dust and synchrotron tracers, we use \Planck\ 353\,GHz, scaled by $\beta=0.039$ and WMAP K band, scaled by $\alpha=0.0095$. The mask used for the foreground template fitting, both for data and simulations, retains 70\,\% of the sky. The other masks are used in the cosmological analysis.

Figure~\ref{fig:spectrum_comparison_2018} shows the $100\times143$ $EE$ power spectrum of the \srolltwo\ maps compared with the \Planck\ 2018 power spectrum both on $50$\% of the sky. The error bars are obtained combining a Monte Carlo of CMB signal with $\tau=0.055$ with \noiseMC, and computing the QML power spectrum from all the maps. The quadrupole affecting \Planck\ 2018 analysis is radically reduced by the new map-making procedure.

\begin{figure}[h]
 \includegraphics[width=0.48\textwidth]{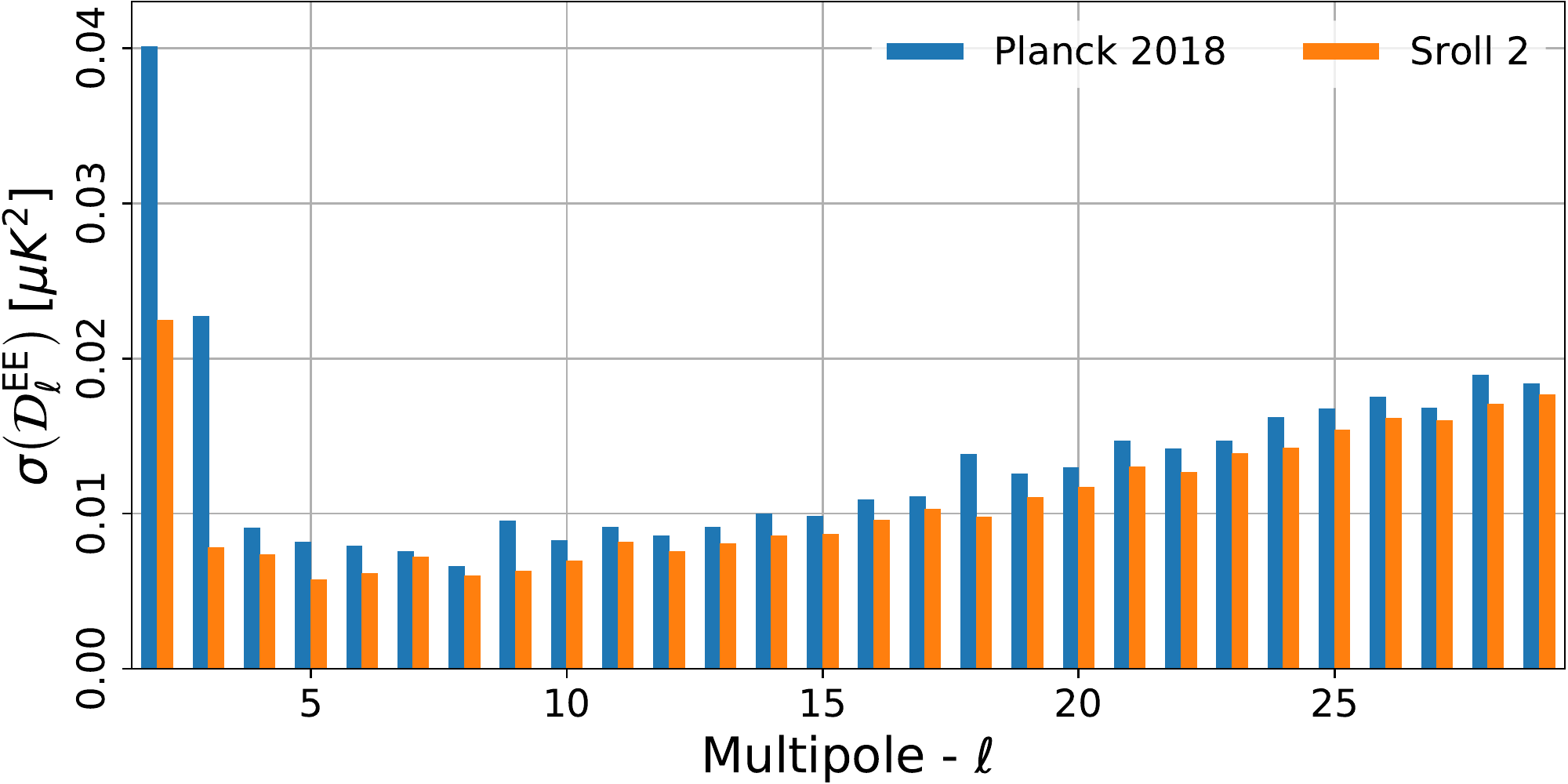}
 \caption{Comparison between \Planck\ 2018 legacy release and \srolltwo\ error bars for $100\times143$ spectrum both on 50\% of the sky. For \srollone\ \Planck\ 2018 FFP10 \cite{planck2016-l03} simulations have been used, for \srolltwo\ \noiseMC\ simulations presented in \citet{Delouis:2019bub}. Cosmic variance here is not included. \label{fig:error_comparison_2018_bware}}
 \end{figure}

\begin{figure}[h!]
 \includegraphics[width=0.48\textwidth]{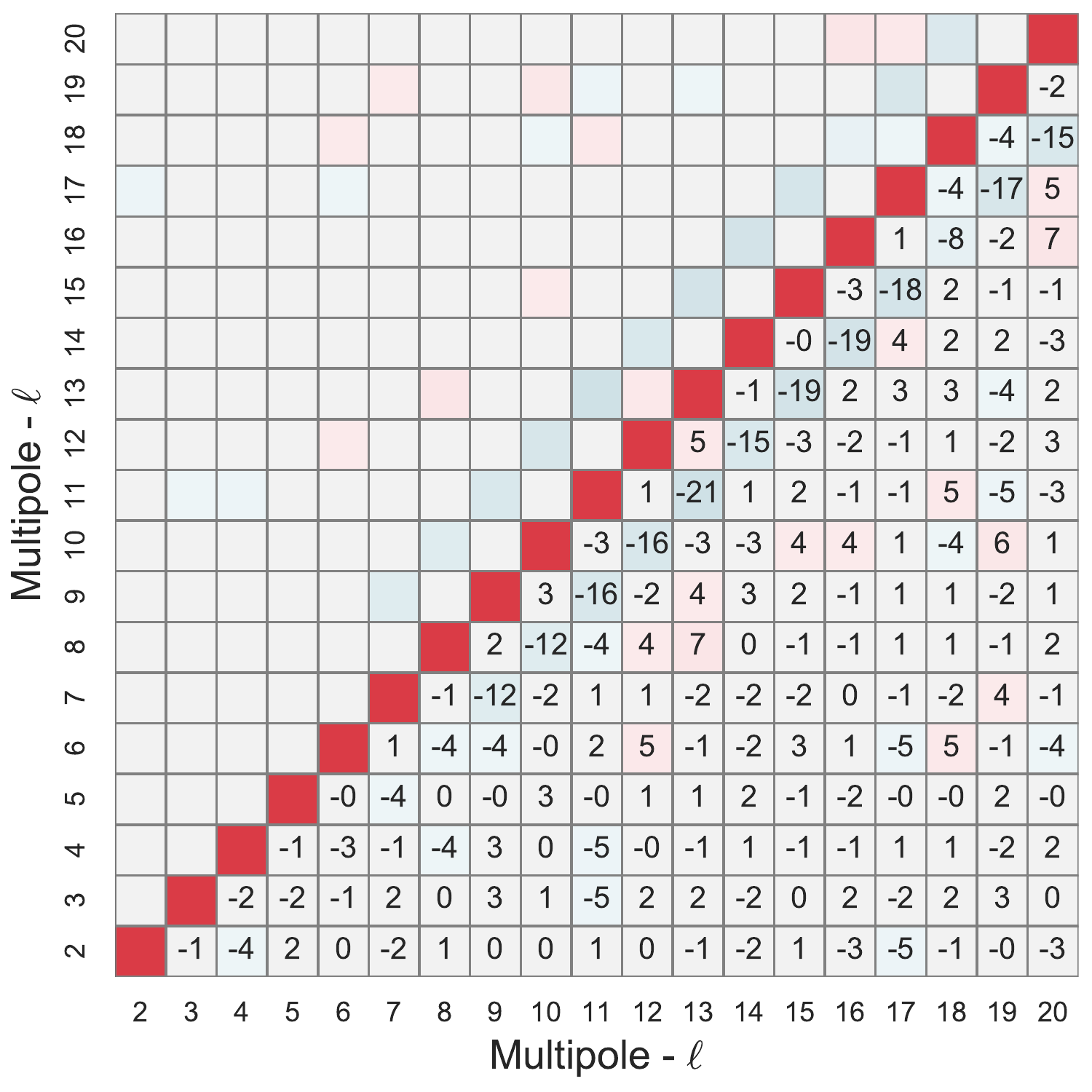}
 \caption{Correlation matrix in $[\%]$ for $EE$ power spectrum below $\ell=20$ estimated from 500 signal (with $\tau=0.055$) + \noiseMC\ simulations. The different multipoles are substantially uncorrelated.\label{fig:correlation_matrix}}
 \end{figure}

In Fig.~\ref{fig:error_comparison_2018_bware}, we compare the error bars purged from cosmic variance obtained in \srolltwo\ maps with the ones of \Planck\ 2018 analysis. With the \srolltwo\ maps, we manage to halve quadrupole and octupole errors with respect to the \Planck\ 2018 analysis. Overall the entire range of multipoles sensitive to reionization optical depth shows a clearly reduced level of residual systematic effects and a lower variance. Furthermore, the $\mathcal{C}_\ell$s are weakly correlated as shown in Fig. \ref{fig:correlation_matrix}. In the range relevant  for $\tau$ estimation $\ell=[2\dots8]$, the multipoles are substantially uncorrelated, with the scatter observed in the off-diagonal correlation perfectly compatible with the number of \noiseMC\ simulations. In the region where the $EE$ signal is expected to be very small in $\Lambda$CDM model ($\ell=[10\dots20]$) we notice the presence of a weak (up to $20\%$) $\ell,\ell+2$ correlation, nevertheless this feature is not expected to affect substantially the $\tau$ estimation.

\begin{figure}[h]
 \includegraphics[width=0.48\textwidth]{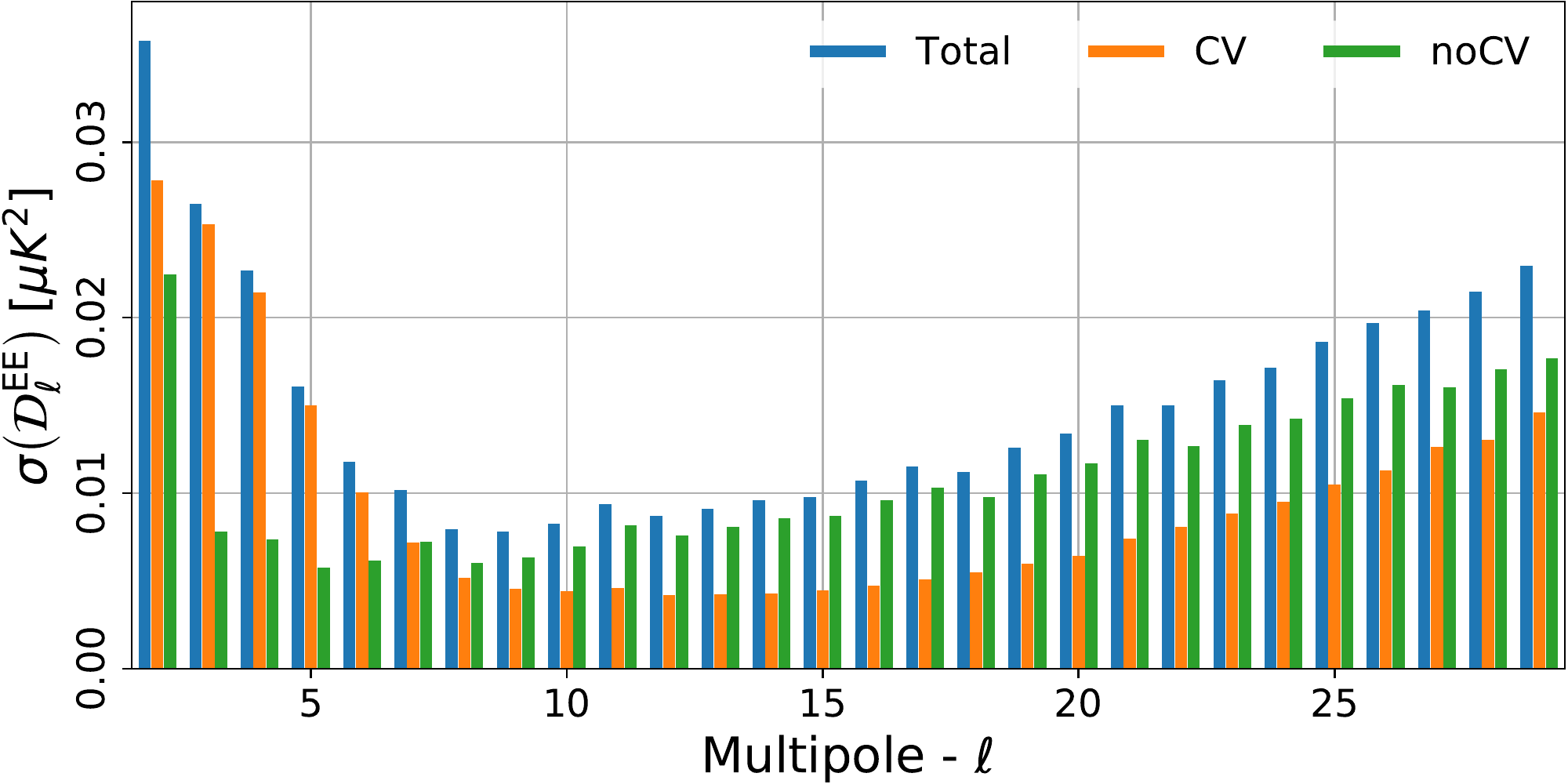}
 \caption{Error comparison for the $100\times143$ spectrum on 60\% of the sky. We show the total error (blue bar), the amount solely due to cosmic variance (orange), and only due to noise and systematic effects (green). The cosmic variance shown corresponds to $\tau=0.055$.\label{fig:error_comparison_bware}}
 \end{figure}

By comparing the fraction of the error due to noise and systematic effects with the cosmic variance for $\tau=0.055$, in the range $\ell=[2\dots6]$, we notice that, for the first time, we start being dominated by the latter, as shown in Fig.\ref{fig:error_comparison_bware}. All the error bars are obtained using simulations and not computed analytically.

\begin{figure}[h!]
 \includegraphics[width=0.48\textwidth]{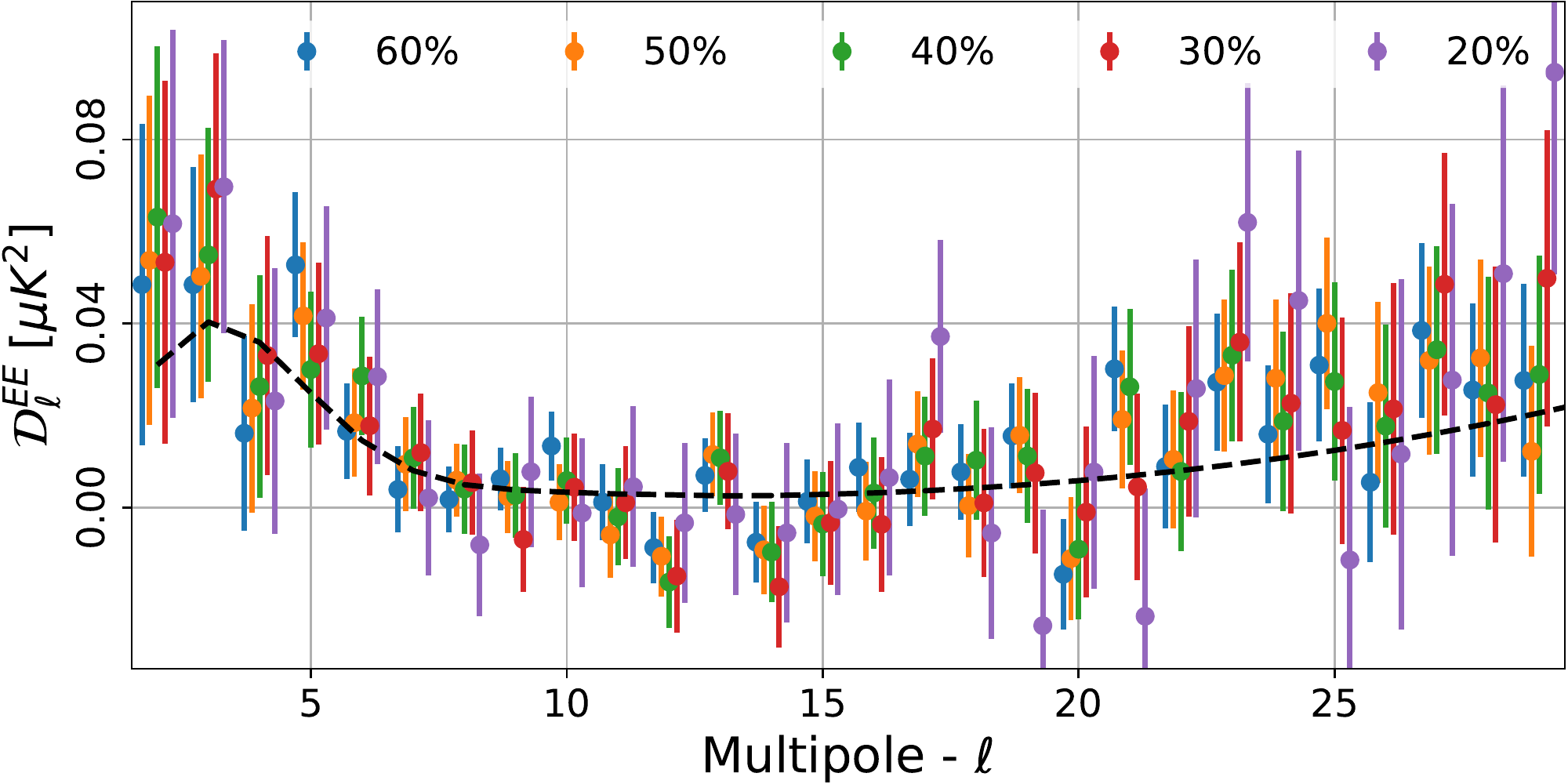}
 \caption{$EE$ power spectra of $100\times143$ for different sky fractions. Error bars are obtained from the distribution of 500 signal (with $\tau=0.055$) + \noiseMC\ simulations. The black solid line corresponds to a $EE$ power spectrum with $\tau=0.055$. \label{fig:spectrum_fsky_comparison}}
 \end{figure}

In Fig.~\ref{fig:spectrum_fsky_comparison}, we compare $EE$ power spectrum obtained with different masks. The multipole $\ell=5$ shows the largest variation throughout the various masks. We verify, using simulations, that this variation is always consistent with $2\,\sigma$ fluctuation. We emphasize again that \noiseMC\ contains noise, systematic effects and residuals of foreground cleaning.

\begin{figure}[h!]
 \includegraphics[width=0.48\textwidth]{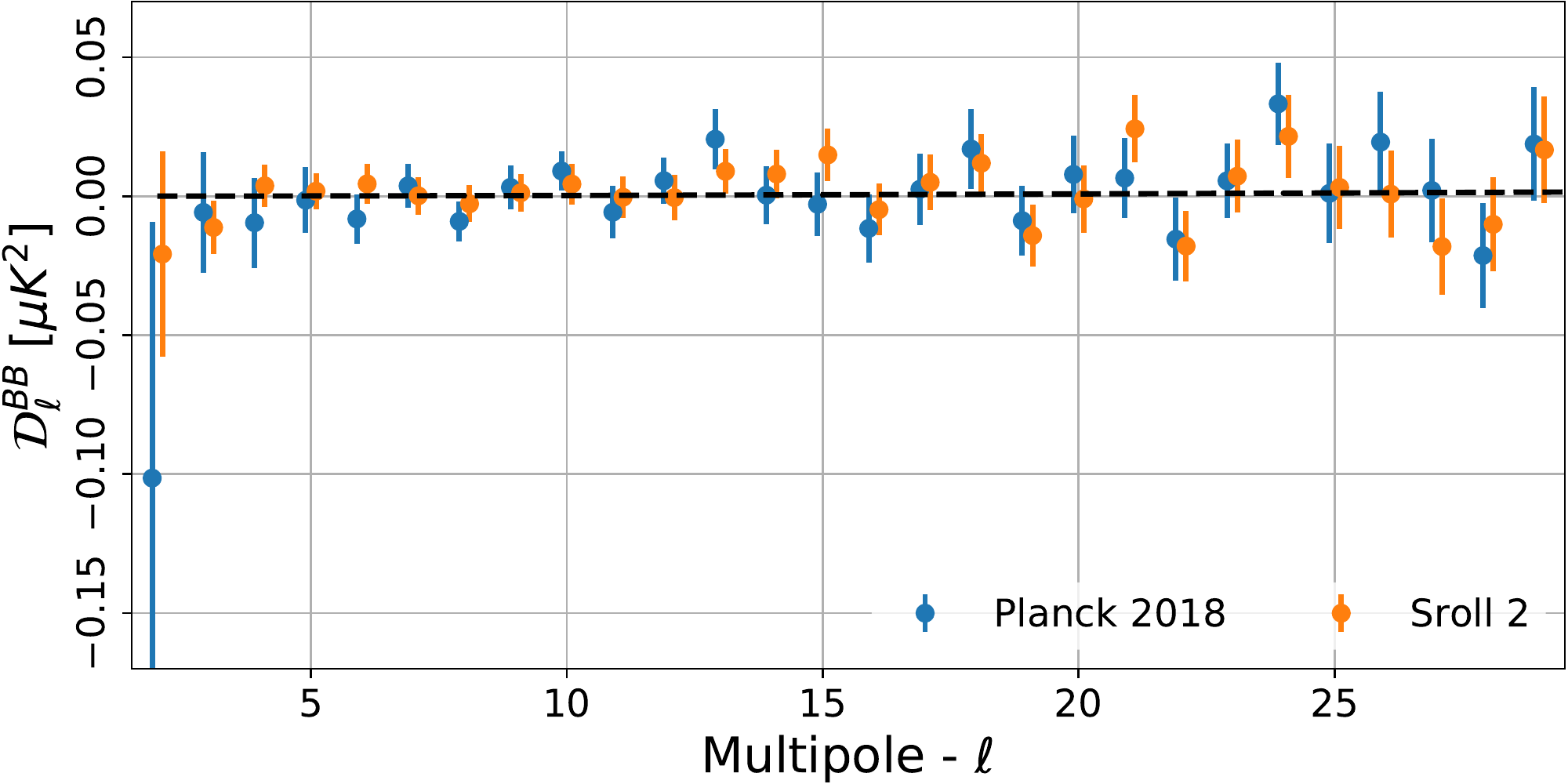}
 \caption{Low-$\ell$ $BB$ cross spectrum $100\times143$ for the \Planck\ 2018 legacy release (blue points) and for the \srolltwo\ maps (orange). The mask used retains 50\% of the sky. The error bars are Monte Carlo based and do include cosmic variance.\label{fig:spectrum_comparison_2018_BB}}
 \end{figure}

As final test, we show in Fig.~\ref{fig:spectrum_comparison_2018_BB} the $BB$ power spectrum obtained from the cross-correlation of 100 and 143\,GHz \srolltwo\ maps. As reference, we jointly plot \Planck\ 2018 legacy release $BB$ power spectrum \citep{planck2016-l05}. Both spectra are compatible with null signal, with \srolltwo\, being more constraining at very large scale. The probability to exceed is PTE=0.73, for  \Planck\ 2018, and PTE=0.86, for \srolltwo. The large negative quadrupole in \Planck\ 2018 legacy release, related to ADCNL residuals (see \citet{planck2016-l03} and \citet{planck2016-l05}), is almost completely reabsorbed in the new analysis.
As final test, assuming the empirical distribution of the \noiseMC\ simulations and the power spectra  measured on data computed on the 50\% mask, in Tab.~\ref{tab:empirical_pte_ellbyell} we report, $\ell$-by-$\ell$, the percentage of simulations that have an absolute value of the difference between $\mathcal{D}_\ell$ and the barycenter of the distribution larger than the same quantity measured on data. Also in this case, the overall agreement is excellent, with no particular outliers.

\begin{table}[h!]
	\begingroup
	\caption{Percentage of signal plus \noiseMC\ simulations that $\ell$-by-$\ell$ have absolute difference between $\mathcal{D}_\ell$ and the mean of the empirical distribution larger than the data.}
	\label{tab:empirical_pte_ellbyell}
	\nointerlineskip
	\vskip -3mm
	\footnotesize
	\setbox\tablebox=\vbox{
		\newdimen\digitwidth
		\setbox0=\hbox{\rm 0}
		\digitwidth=\wd0
		\catcode`*=\active
		\def*{\kern\digitwidth}
		\newdimen\signwidth
		\setbox0=\hbox{+}
		\signwidth=\wd0
		\catcode`!=\active
		\def!{\kern\signwidth}
		\halign{\hbox to 0.8in{#\leaderfil}\tabskip=1em&
			\hfil#\hfil\tabskip=10pt&
			\hfil#\hfil\tabskip=10pt&
			\hfil#\hfil\tabskip=10pt&
			\hfil#\hfil\tabskip=10pt&
			\hfil#\hfil\tabskip=10pt&
			\hfil#\hfil\tabskip=0pt\cr
			\noalign{\doubleline}
			\noalign{\vskip 3pt}
			\omit\hfil Mulitpole\hfil  & TE & EE & BB & TB & EB\cr
			\noalign{\vskip 3pt\hrule\vskip 5pt}
2 & 58.1 & 38.0 & 76.5 & 87.5 & 56.2\cr
3 & 72.2 & 20.4 & 6.8 & 42.8 & 90.9\cr
4 & 52.8 & 62.5 & 50.1 & 88.5 & 53.9\cr
5 & 28.7 & 93.4 & 52.7 & 23.6 & 27.7\cr
6 & 77.7 & 67.1 & 85.4 & 48.2 & 99.8\cr
7 & 93.1 & 91.4 & 35.9 & 68.8 & 72.7\cr
8 & 94.2 & 55.4 & 56.0 & 22.7 & 35.5\cr
9 & 99.4 & 83.2 & 12.8 & 43.8 & 93.2\cr
10 & 75.0 & 59.1 & 99.3 & 52.1 & 5.6\cr
11 & 16.3 & 62.6 & 46.9 & 37.0 & 52.5\cr
12 & 13.9 & 76.8 & 74.1 & 67.2 & 23.7\cr
13 & 24.5 & 32.0 & 95.7 & 91.1 & 39.9\cr
14 & 16.9 & 32.4 & 83.5 & 1.8 & 61.8\cr
15 & 52.1 & 12.8 & 98.6 & 2.9 & 7.3\cr
16 & 71.5 & 45.4 & 87.0 & 78.9 & 65.7\cr
17 & 38.4 & 75.6 & 89.4 & 86.8 & 59.3\cr
18 & 61.4 & 42.3 & 4.1 & 5.9 & 2.4\cr
19 & 32.3 & 13.7 & 11.9 & 84.5 & 14.4\cr
20 & 25.6 & 81.2 & 97.5 & 52.3 & 24.0\cr
21 & 43.4 & 8.0 & 70.6 & 31.6 & 51.0\cr
22 & 92.5 & 13.4 & 24.7 & 84.6 & 64.0\cr
23 & 34.5 & 68.8 & 91.5 & 22.4 & 35.6\cr
24 & 28.8 & 22.3 & 6.1 & 51.9 & 33.6\cr
25 & 17.5 & 93.8 & 40.9 & 15.8 & 57.6\cr
26 & 55.4 & 89.1 & 15.1 & 88.1 & 14.5\cr
27 & 55.5 & 21.0 & 84.7 & 3.2 & 26.1\cr
28 & 54.6 & 44.4 & 6.4 & 36.2 & 25.2\cr
29 & 76.9 & 57.4 & 86.5 & 88.3 & 91.8\cr
	\noalign{\vskip 5pt\hrule\vskip 3pt}}}
	\endPlancktable
	\endgroup
\end{table}

\section{Likelihood and $\tau$ estimation}\label{sec:likelihood}
 
Following the procedure presented in \citet{planck2016-l05}, we build a likelihood for $\tau$, based on $100\times143$ $EE$ power spectrum in the multipole range $2-29$. In details:
\begin{itemize}
\item we generate 101 theoretical power spectra, $\mathcal{C}_\ell^{\rm th}\left(\tau,\boldsymbol{\theta}\right)$, equally spaced in the range $\tau=\left[0,0.1\right]$, varying accordingly $A_s$ such that $10^{9} A_s\, e^{-2\tau}=1.875$ as in \citet{planck2014-a10}. The other $\Lambda$CDM cosmological parameters ($\boldsymbol{\theta}$) are fixed to the best fit model of \citet{planck2016-l06};
\item for each $\mathcal{C}_\ell^{\rm th}\left(\tau,\boldsymbol{\theta}\right)$, we build a CMB signal Monte Carlo of 500 maps;
\item we combine \noiseMC\ with the signal Monte Carlo, and we compute the $100\times143$ $EE$ spectrum for each realization. We compute the power spectrum also for data, $\mathcal{C}_\ell^{\rm data}$;  
\item histogramming the simulations $\ell$-by-$\ell$ and $\tau$-by-$\tau$, we build empirically the probability $\mathcal{P}\left(\mathcal{C}_\ell|\tau ;\boldsymbol{\theta} \right)$;
\item with a piecewise polynomial function $f_\ell\left(\mathcal{C}_\ell;\tau,\boldsymbol{\theta} \right)$ we interpolate $\ln\mathcal{P}\left(\mathcal{C}_\ell|\tau ;\boldsymbol{\theta} \right)$ in order to smooth the scatter due to limited number of simulations available;
\item assuming negligible correlation between multipoles (see Fig.~\ref{fig:correlation_matrix}), we compute the probability for each $\tau$ value in our grid:
\be
\mathcal{P}\left({\rm data} |\tau;\boldsymbol{\theta} \right)= \exp\left(\sum_{\ell=2}^{29} f_\ell\left(\mathcal{C}_\ell^{\rm data};\tau,\boldsymbol{\theta} \right)\right). \label{eq:sum_posterior}
\ee
\end{itemize}

With this algorithm, we can draw slices of probability for $\tau$ adopting different sky fractions and multipole ranges, in order to stress the stability of the analysis and to perform consistency tests.

\begin{figure}[h]
 \includegraphics[width=0.48\textwidth]{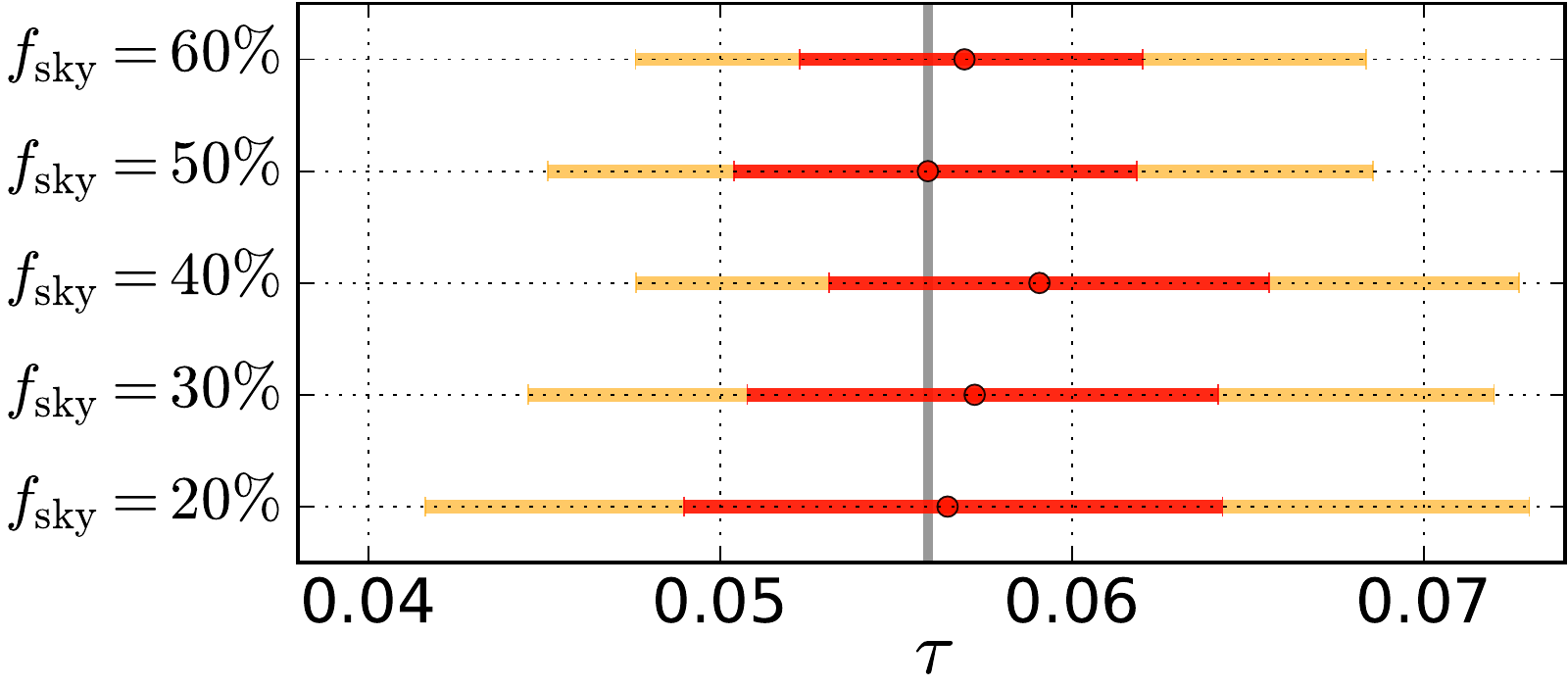}
 \caption{Values of $\tau$ obtained varying the sky fraction used for power spectrum estimation of $100\times143$. In this and the following plots the round points represent best-fit values and, red and yellow bars 68\% and 95\% C.L. respectively. \label{fig:plot_tau_fsky}}
 \end{figure}

As a first consistency check, we test how the sky fraction used to compute the power spectrum impacts the $\tau$ constraint. We explore the same masks used in Fig.~\ref{fig:spectrum_fsky_comparison} and shown in Fig~\ref{fig:masks}. In Fig.~\ref{fig:plot_tau_fsky}, we show a whisker plot with best-fit values, 68\% and 95\% confidence levels on $\tau$ for spectra computed with different QML masks, ranging from 20 to 60\% of used sky. 
The $\tau$ posteriors are stable, all within one $\sigma$. We verify on simulations the statistical consistency of the $\tau$ values computed on different masks, finding a consistency always better than $1.3\,\sigma$ throughout the various masks, having accounted for the common sky, noise and systematics. As baseline, we adopt the $50\%$ mask where we measure a reionization optical depth of:
\be
\tau=0.0566_{-0.0062}^{+0.0053} \qquad (68\%,\textrm{   Sroll2 $EE$ spectrum}),\label{eq:tau_result}
\ee

\noindent having fixed $10^{9} A_s\, e^{-2\tau}=1.875$. 

In the following part of this section, we show tests performed only on the $50\%$ sky mask.

\begin{figure}[h]
 \includegraphics[width=0.48\textwidth]{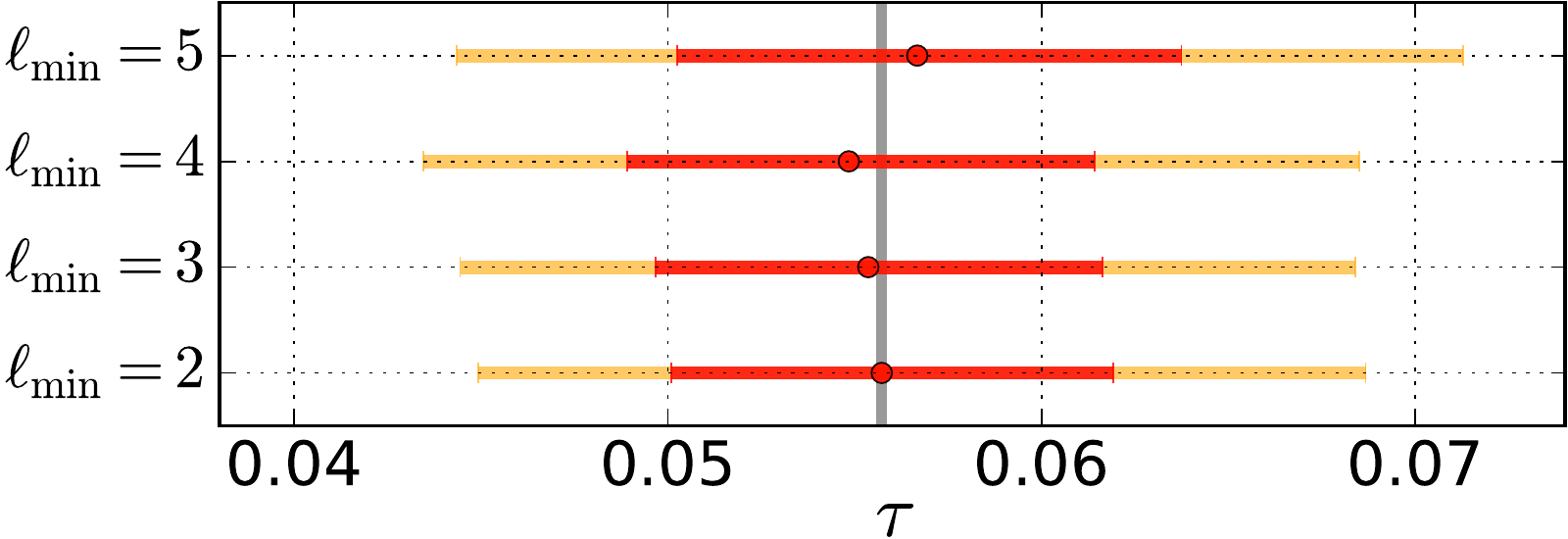}
 \caption{Values of $\tau$ obtained changing the minimum multipole used in the likelihood code.\label{fig:plot_tau_ellmin}}
 \end{figure}

Figure~\ref{fig:plot_tau_ellmin} shows the effect of changing the minimum multipole used in Eq.~\ref{eq:sum_posterior}. The $\tau$ posteriors are stable up to $\ell_{\rm min}=5$, further explorations being less meaningful due to the drop of the reionization feature above those multipoles.

\begin{figure}[h]
 \includegraphics[width=0.48\textwidth]{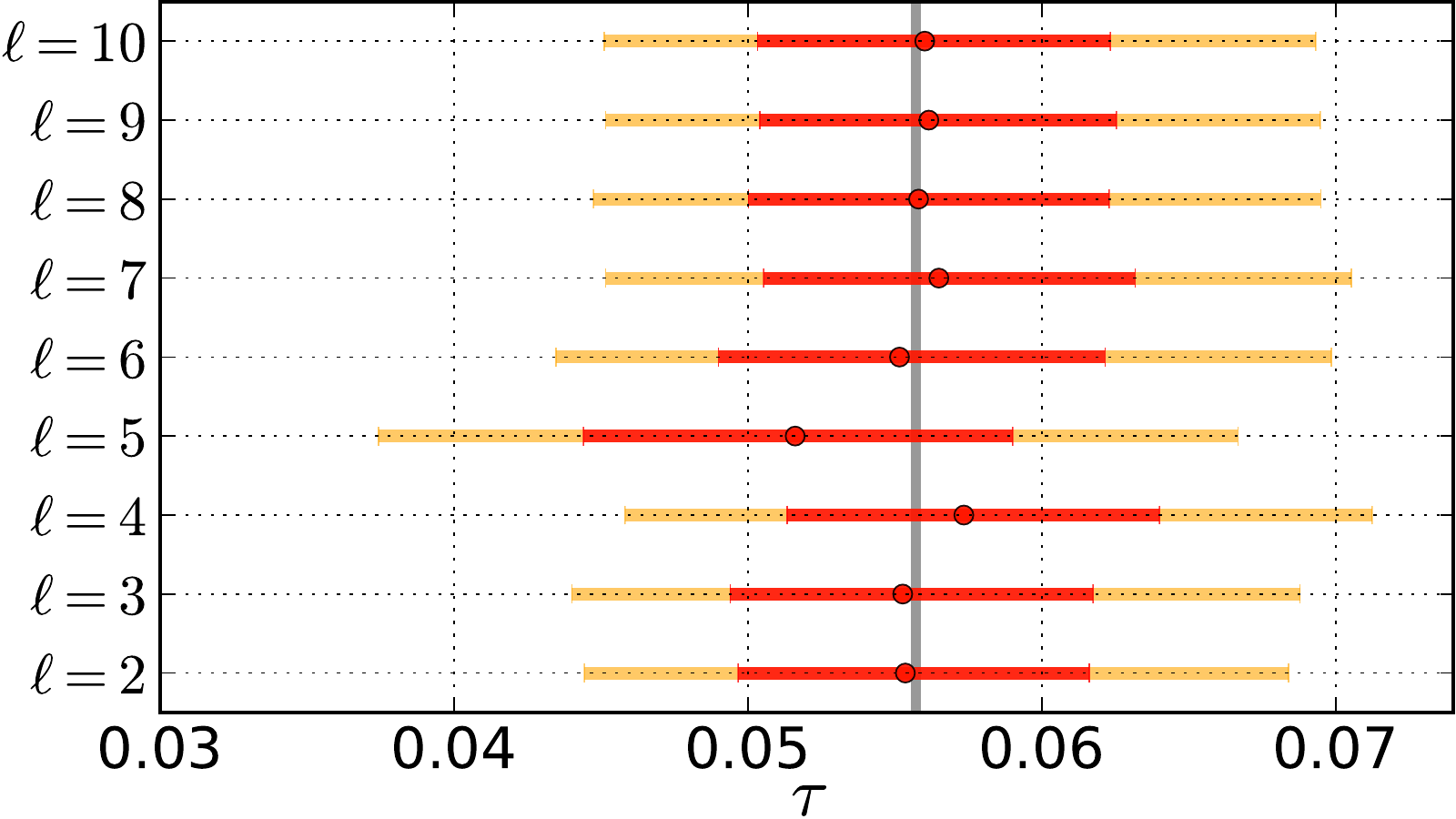}
 \caption{Posteriors of $\tau$ obtained removing one multipole at a time.\label{fig:plot_tau_removedell}}
 \end{figure}

In Fig.~\ref{fig:plot_tau_removedell}, we test the stability of $\tau$ posterior when one multipole at a time is removed from the summation in Eq.~\ref{eq:sum_posterior}. The maximum variation is observed when $\ell=5$ is removed causing a roughly half-$\sigma$ shift in the $\tau$ posterior. This shift was consistently observed by analogous analysis performed on previous versions of the same HFI data (see e.g. \citet{planck2014-a10} figure D.9 and \citet{planck2016-l05} figure 14) with \srolltwo\ being less discrepant despite the smaller overall error budget. Also in this case, we verify on simulations that the $\tau$ obtained removing $\ell=5$ is consistent with a $1.2\,\sigma$ statistical fluctuation when compared with the one obtained using the full range. 

\begin{figure}[h]
 \includegraphics[width=0.48\textwidth]{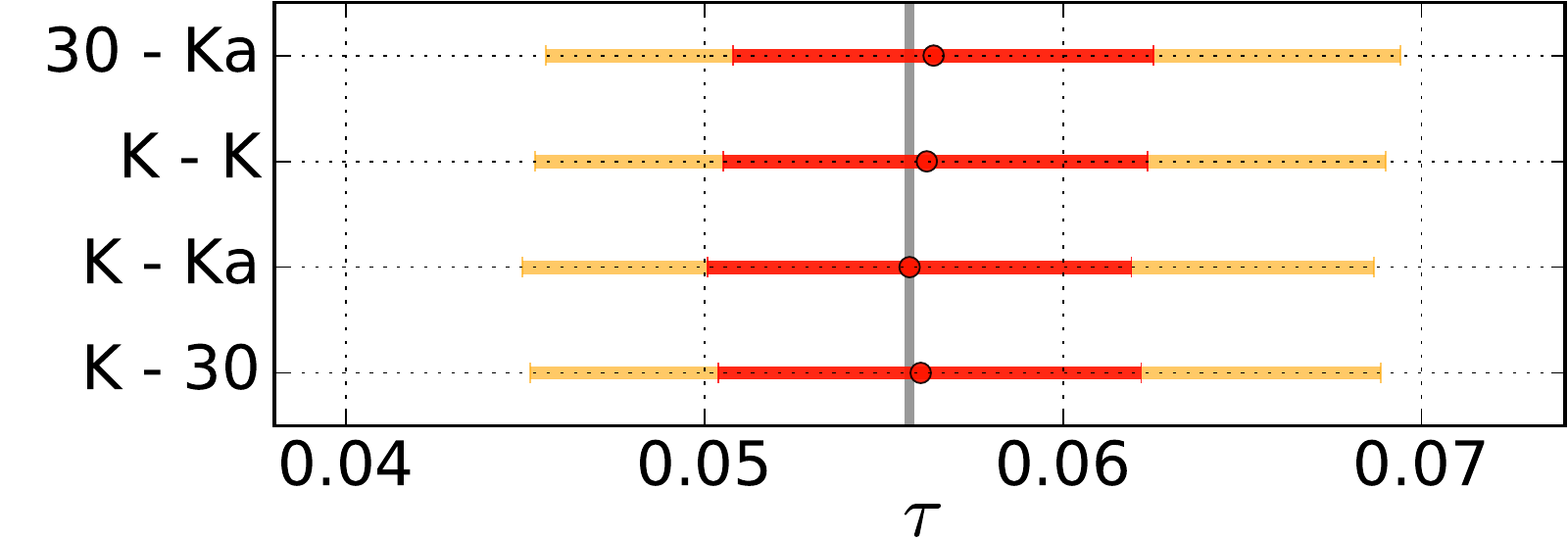}
 \caption{Posteriors of $\tau$ obtained using different synchrotron tracers for 100 and 143\,GHz. The channels reported on the left side of the figure refer to the synchrotron tracers used for 100 and 143\,GHz respectively.\label{fig:plot_tau_sync_tracer}}
 \end{figure}

We also explore the stability of the $\tau$ constraint changing  the synchrotron tracers used respectively for 100 and 143\,GHz. In figure~\ref{fig:plot_tau_sync_tracer}  we show $\tau$ posteriors obtained using different combinations of the available synchrotron tracers, WMAP K band, WMAP Ka band or LFI 30~GHz \citep{planck2016-l02}, all the posteriors are extremely consistent demonstrating that the synchrotron subtraction does not represent a critical point, as already discussed in \citet{planck2016-l05}. 

\begin{figure}[h]
 \includegraphics[width=0.48\textwidth]{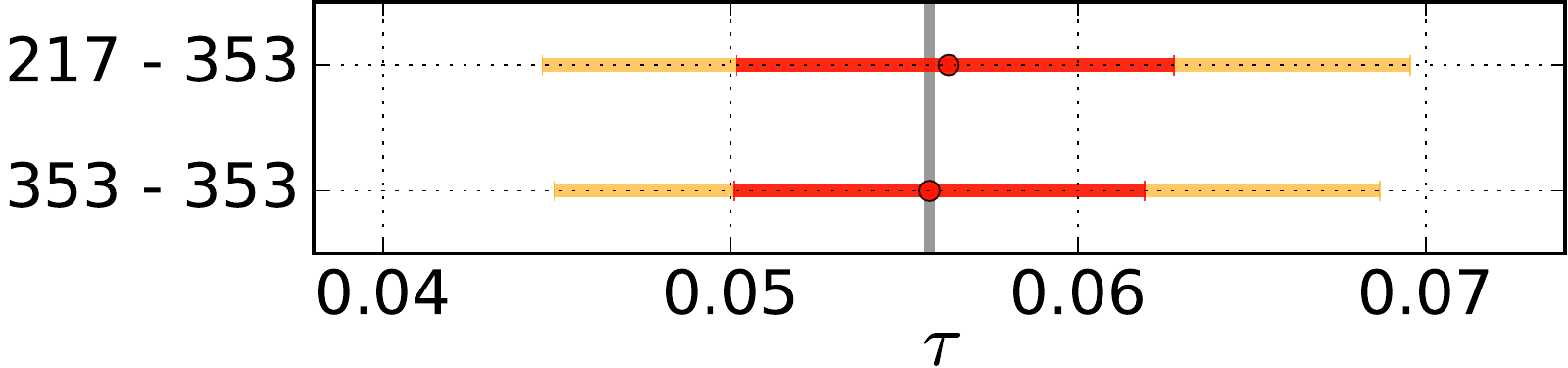}
 \caption{Posteriors of $\tau$ obtained using different dust tracers for 100\,GHz. Similarly to what show in Fig.~\ref{fig:plot_tau_sync_tracer} the channels reported in the y-axis label refer to the dust tracers used respectively for 100 and 143\,GHz. \label{fig:plot_tau_dust_tracer}}
 \end{figure}

We tested the quality of the dust removal employing the 217\,GHz instead of 353\,GHz for the cleaning of 100\,GHz. In figure~\ref{fig:plot_tau_dust_tracer} we compare the $\tau$ posterior obtained cleaning both 100 and 143\,GHz using 353\,GHz with the one obtained by cleaning 100\,GHz with 217\,GHz and 143\,GHz with 353\,GHz. The consistency is remarkable, with only the latter showing slightly larger error bars, likely due to the smaller leverage of 217 \,GHz  not fully compensated by the lower noise.

 Finally, we attempt a similar analysis on the $TE$ spectrum only, measuring $\tau$ for the same 50\% mask, finding:

 \be
\tau=0.057_{-0.013}^{+0.012} \qquad (68\%,\textrm{   Sroll2 $TE$ spectrum}),\label{eq:tau_result_TE}
\ee

\noindent which nicely confirms the $EE$ based result. It is worth to mention here that the poor PTEs for the null TE spectra found in the \Planck\ 2018 likelihood analysis \citep{planck2016-l05} are still present in this version of the data, even if with slightly less significance. We recall also that we employ the same \commander\ solution used in the \Planck\ 2018 likelihood, thus based on \srollone\ maps.

\section{Impact on cosmology}\label{sec:cosmology}

Following the method presented in \citet{planck2016-l05} and used for the \planck\ \simall\ likelihood (i.e. \lowE\ in \planck\ 2018 legacy release) we build a likelihood for \srolltwo\ 100x143 $EE$ power spectrum\footnote{For details about validation and performances of the likelihood approximation see \citet{Gerbino:2019okg}.}. We call this new likelihood module \lowEStwo\ \footnote{The likelihood module is built within the \clik\ infrastructure \citep{planck2013-p08,planck2013-p28,planck2014-ES,planck2016-ES} and it is available on \href{http://sroll20.ias.u-psud.fr}{http://sroll20.ias.u-psud.fr} or on \href{https://web.fe.infn.it/~pagano/low\_ell\_datasets/sroll2}{https://web.fe.infn.it/$\sim$pagano/low\_ell\_datasets/sroll2} }.

Superseeding the \planck\ \lowE\ likelihood, we combine the \lowEStwo\ with the high-$\ell$ \plik\ 2018 likelihood and with the \commander\ 2018 low-$\ell$ temperature likelihood in order to constrain cosmological parameters. In this section, we explore the cosmological parameter space making use of the \cosmomc\footnote{\url{http://cosmologist.info/cosmomc}} package \citep{Lewis:2002ah} based on \camb\footnote{\url{http://camb.info}} Boltzmann code \citep{Lewis:1999bs}. The global settings in terms of parametrization and assumptions are coherent with \citet{planck2016-l06}.

 First of all, we combine  \lowEStwo\ with only with \commander\ 2018 temperature likelihood, and we estimate the cosmological parameters only sampling $\ln(10^{10} A_\mathrm{s})$ and $\tau$, keeping fixed the other parameters to \Planck\ TT+\lowE\ bestfits, measuring 
  
\be
\tau=0.0579_{-0.0067}^{+0.0056} \qquad (68\%,\textrm{  \commander\ TT+}\text{\lowEStwo}),\label{eq:tau_commanderTT_plus_lowEE}
\ee

which is directly comparable with the bounds shown in Eq.~\ref{eq:past_tau_values}. The amplitude of the scalar perturbations preferred by the temperature likelihood is substantially low (see e.g. \citet{planck2016-l05} Table 4 and 12) which is compensated by an increase of the $\tau$ value. 
Opening the other $\Lambda$CDM parameters, and adding the TT likelihood drives the value of $\tau$ upwards

\be
\tau=0.0575_{-0.0069}^{+0.0056} \qquad (68\%,\textrm{   TT+}\text{\lowEStwo}),\label{eq:tau_TT_plus_lowEE}.
\ee

Similar behaviour was also observed in \citet{planck2016-l06} and \citet{planck2014-a15} and is mainly due to the  $A_s\, e^{-2\tau}$ degeneracy broken by the high-$\ell$ lensing in the temperature spectrum.
 The addition of high-$\ell$ polarization drives again upward $A_s$ and thus the optical depth up to:

\be
\tau=0.0591_{-0.0068}^{+0.0054} \qquad (68\%,\textrm{   TT,TE,EE+}\text{\lowEStwo}).\label{eq:tau_TTTEEE_plus_lowEE}
\ee

The fluctuation amplitude can be directly constrained at late times by CMB lensing reconstruction power spectrum \citep{planck2016-l08}, partially degenerated with the matter density, while the BAO measurements constrain very efficiently the geometry of the late universe (see \citet{planck2016-l06} for more details on those datasets). Nonetheless the combination of \Planck\ 2018 lensing likelihood and BAO measurements with the primary CMB anisotropies does not improve significantly the $\tau$ constraint:

\be
\tau=0.0599_{-0.0064}^{+0.0054} \, \onesig{\textrm{   TT,TE,EE+\lowEStwo+Lensing+BAO}}. \label{eq:tau_TTTEEE_plus_lowEE_plus_lensing_plus_BAO}
\ee

Assuming a $\tanh$ parametrization of the ionization fraction, the $\tau$ constrain can be translated into a mid-point redshift of reionization of:

\be
z_{\mathrm{re}}=8.21\pm 0.58 \, \onesig{\textrm{   TT,TE,EE+\lowEStwo}+Lensing+BAO}, \label{eq:tau_TTTEEE_plus_lowEE_plus_lensing_plus_BAO}
\ee

consistent with the latest astrophysical constraint of high-redshift quasars (see e.g. \citet{Becker:2001ee}, \citet{Fan:2005es} and \citet{Bouwens:2015vha} for an exhaustive comparison).

The combination of low and high-$\ell$ likelihoods breaks efficiently the $A_s\, e^{-2\tau}$ degeneracy, giving:

\begin{multline}
\ln(10^{10} A_\mathrm{s})=3.054\pm0.012 \\ \onesig{\textrm{   TT,TE,EE+\lowEStwo}+Lensing+BAO}.\label{eq:lnAs_TTTEEE_plus_lowEE_plus_lensing_plus_BAO}
\end{multline}

In the context of $\Lambda$CDM model, this bound can be directly translated in the $\sigma_8$ parameter

\begin{multline}
\sigma_8=0.8128\pm0.0053 \\ \onesig{\textrm{   TT,TE,EE+\lowEStwo}+Lensing+BAO},\label{eq:s8_TTTEEE_plus_lowEE_plus_lensing_plus_BAO}
\end{multline}

which measures the amplitude of the matter power spectrum on the scale of $8h^{-1}\rm{Mpc}$.

Bounds on the $\Lambda$CDM native parameters and some meaningful derived ones are reported in Tab.~\ref{tab:res} where we compare the results obtained with the \planck\ 2018 baseline likelihood with the ones obtained replacing \lowE\ with \lowEStwo.

\begin{table*}
\begingroup
\caption{Parameter constraints for the base \LCDM\ cosmology \citep[as defined in][]{planck2013-p11}, illustrating the impact of replacing the \lowE\ likelihood  with the \lowEStwo\ likelihood presented in the paper. We also show the change when including the high-$\ell$ polarization.}
\label{tab:res}
\openup 5pt
\newdimen\tblskip \tblskip=5pt
\nointerlineskip
\vskip -3mm
\normalsize
\setbox\tablebox=\vbox{
    \newdimen\digitwidth
    \setbox0=\hbox{\rm 0}
    \digitwidth=\wd0
    \catcode`"=\active
    \def"{\kern\digitwidth}
    \newdimen\signwidth
    \setbox0=\hbox{+}
    \signwidth=\wd0
    \catcode`!=\active
    \def!{\kern\signwidth}
\halign{
     \hbox to 0.9in{$#$\leaderfil}\tabskip=1.5em&
     \hfil$#$\hfil&
     \hfil$#$\hfil&\hfil$#$\hfil&
     \hfil$#$\hfil\tabskip=0pt\cr
\noalign{\doubleline}
\multispan1\hfil \hfil&\multispan1\hfil TT+\lowE\hfil&\multispan1\hfil TT+\lowEStwo\hfil&\multispan1\hfil TTTEEE+\lowE\hfil&\multispan1\hfil TTTEEE+\lowEStwo\hfil\cr
\noalign{\vskip -3pt}
\omit\hfil Parameter\hfil&\omit\hfil 68\,\% limits\hfil&\omit\hfil 68\,\% limits\hfil&\omit\hfil 68\,\% limits\hfil&\omit\hfil 68\,\% limits\hfil\cr
\noalign{\vskip 3pt\hrule\vskip 5pt}
\Omega_{\mathrm{b}} h^2&0.02212\pm 0.00022&0.02214\pm 0.00021&0.02236\pm 0.00015&0.02237\pm 0.00015\cr
\Omega_{\mathrm{c}} h^2&0.1206\pm 0.0021&0.1205\pm 0.0021&0.1202\pm 0.0014&0.1201\pm 0.0013\cr
100\theta_{\mathrm{MC}}&1.04077\pm 0.00047&1.04080\pm 0.00047&1.04090\pm 0.00031&1.04090\pm 0.00031\cr
\tau&0.0522\pm 0.0080&0.0574_{-0.0069}^{+0.0056}&0.0544^{+0.0070}_{-0.0081}&0.0591_{-0.0068}^{+0.0054}\cr
\ln(10^{10} A_\mathrm{s})&3.040\pm 0.016&3.051\pm 0.013&3.045\pm 0.016&3.054\pm 0.013\cr
n_\mathrm{s}&0.9626\pm 0.0057&0.9631\pm 0.0056&0.9649\pm 0.0044&0.9651\pm 0.0043\cr
\noalign{\vskip 3pt\hrule\vskip 5pt}
H_0&66.88\pm 0.92&66.95\pm 0.90&67.27\pm 0.60&67.32\pm 0.60\cr
\Omega_{\mathrm{m}}&0.321\pm 0.013&0.320\pm 0.013&0.3166\pm 0.0084&0.3158\pm 0.0082\cr
\Omega_{\mathrm{\Lambda}}&0.679\pm 0.013&0.680\pm 0.013&0.6834\pm 0.0084&0.6842\pm 0.0082\cr
\sigma_8&0.8118\pm 0.0089&0.8155\pm 0.0083&0.8120\pm 0.0073&0.8154\pm 0.0067\cr
z_{\mathrm{re}}&7.50\pm0.82&8.04\pm 0.60&7.68\pm0.79&8.14\pm 0.60\cr
10^9 A_{\mathrm{s}} &2.092\pm 0.034&2.113\pm 0.028&2.101^{+0.031}_{-0.034}&2.120\pm 0.028\cr
10^9 A_{\mathrm{s}} e^{-2\tau}&1.884\pm 0.014&1.884\pm 0.014&1.884\pm 0.012&1.884\pm 0.012\cr
\mathrm{Age}/\mathrm{Gyr}&13.830\pm 0.037&13.827\pm 0.036&13.800\pm 0.024&13.798\pm 0.024\cr
\noalign{\vskip 5pt\hrule\vskip 3pt}
} 
} 
\endPlancktable
\endgroup
\end{table*}

We also consider minimal one parameter extensions of the  $\Lambda$CDM model such as  $\Omega_K$, $\Sigma m_\nu$, $N_{eff}$ and $Y_{He}$ finding no relevant changes with respect to the \planck\ 2018 legacy release bounds \citep{planck2016-l06} which reinforce the  overall stability of the \Planck\ 2018 results. This is likely to be connected to the mostly unchanged upper limit on $\tau$, i.e. $\tau \lsim0.07$ at 95\% C. L..

Finally, we explore the phenomenological parameter $A_{L}$ which rescales the lensing potential with respect to the theoretical expectation within $\Lambda$CDM model. Consistently throughout \planck\ releases the CMB power spectra show a preference for $A_{L}>1$ \citep{planck2013-p11,planck2014-a15,planck2016-l06}, see \citet{planck2016-LI} for extensive discussion. Such values of $A_{L}$ are in slight tension with the theoretical expectations and with the ones extracted from the lensing reconstruction power spectrum \citep{planck2016-l08}. Combining temperature and polarization data, in the \planck\ 2018 legacy release, $A_{L}=1.180\pm0.065$ was measured. Replacing \lowE\ with \lowEStwo\ slightly reduces the lensing amplitude down to $A_{L}=1.163\pm0.064$, without changing the overall conclusions of \citet{planck2016-l06}. This is again explainable with the increase of $A_{\mathrm{s}}$ connected with the increase of $\tau$ which allows a slightly lower lensing amplitude. 

\section{Conclusions}

In this paper we present an improved constraint on the reionization optical depth $\tau$, obtained analyzing the \Planck\ HFI data with an updated version of the \sroll\ map-making algorithm  called \srolltwo\ specifically designed to reduce the residual large scale polarization systematics still present in the \Planck\ HFI 2018 legacy maps. Details and performances of the \srolltwo\ algorithm are described extensively in \citet{Delouis:2019bub}. 

The level of residual systematics associated to the first multipoles, relevant for $\tau$ estimation, is brought below the noise level and for the first time the cosmic variance becomes the main source of uncertainty in CMB large scale polarization parameter estimation.

As explained in \citet{planck2016-l05} (see in particular section 2.4) the level of T to P leakage in the \Planck\ 2018 legacy release maps  forced the \Planck\ Collaboration to adopt a strategy for the large scale polarization likelihood entirely based on simulations. Furthermore the difficulty of building reliable covariance matrices leads to use a simulation based likelihood, built on the $EE$ cross spectrum of 100 and 143 \GHz. In the present analysis, we follow the same overall strategy, although the lower level of systematics could have allowed a semi-analytical approach (see e.g. \citet{Mangilli:2015xya,Vanneste:2018azc,Hamimeche:2008ai,Gerbino:2019okg}) which we leave to future analysis. With this method, we measure $\tau=0.0566_{-0.0062}^{+0.0053}$ at $68\%$ C.L. when all the other $\Lambda$CDM parameter are kept fixed.

The main difference with respect to the \planck\ 2018 analysis (which yields $\tau=0.051\pm 0.009$) is based on the correction of the second-order ADCNL effect presented in \citet{Delouis:2019bub} which drastically reduces the dipole and foreground signals distortion allowing to recover almost completely $\ell=2$ and $\ell=3$ for the $\tau$ determination, suppressed in previous analysis by a large variance (see e.g. \citet{planck2014-a10,planck2016-l05,planck2014-a25}). As consequence of this in the \srolltwo\ $EE$ 100x143 spectrum the variance associated to systematics becomes smaller than the noise and cosmic variance making less critical the accuracy of the ADCNL simulation produced. Those aspects together with an improved version of the foreground model causes a $\sim$1-$\sigma$ upward shift in the $\tau$ posterior.

\begin{figure}[h]
 \includegraphics[width=0.48\textwidth]{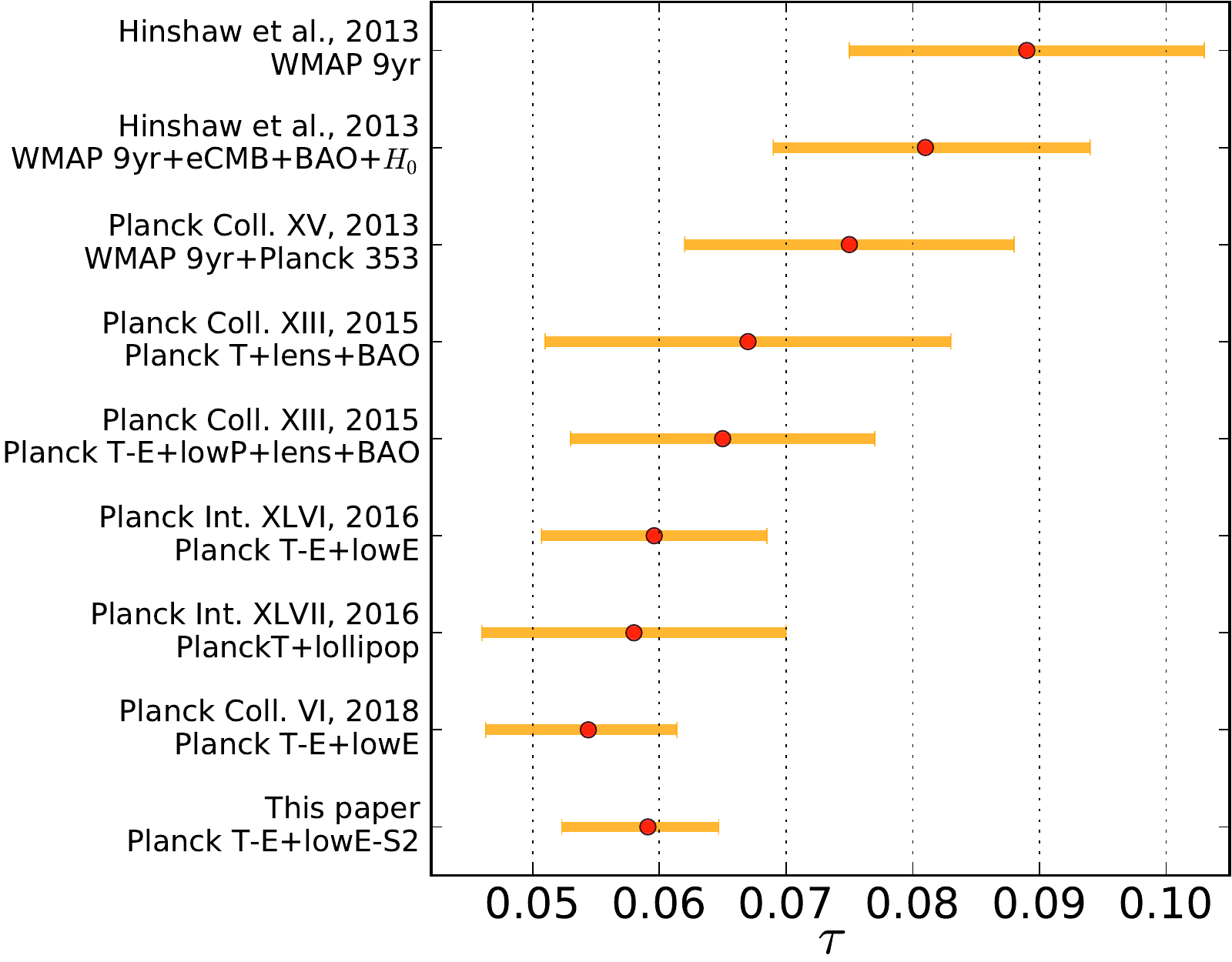}
 \caption{History of $\tau$ determination from \wmap\ to \planck. With Planck T tag we refer to \Planck\ low-$\ell$ and high-$\ell$ temperature likelihood, with Planck T-E, we refer to low-$\ell$ and high-$\ell$ temperature likelihood combined with high-$\ell$ $TE$ and $EE$ likelihood. WMAP 9yr + Planck 353 refers to the WMAP 9yr low-$\ell$ and high-$\ell$ likelihoods with the large scale polarization data cleaned by \planck\ 353GHz.}\label{fig:tau_history}
 \end{figure}

In a more complete parameter exploration, combining the \srolltwo\ likelihood with the temperature and high-$\ell$ polarization likelihood, we measure $\tau=0.059\pm0.006$ at $68\%$ C.L. which represents the strongest constrain on the reionization optical depth to date. The most recent optical depth measurement from CMB data in the context of $\Lambda$CDM model are reported in Fig.~\ref{fig:tau_history}.

Assuming instantaneous reionization, this corresponds to $z_{\rm re}=8.14\pm0.61$ at $68\%$ C.L. The tight bound on $\tau$ breaks efficiently the $A_s\, e^{-2\tau}$ degeneracy reducing the constraint on the fluctuation amplitude down to $\sigma_8=0.8128\pm0.0053$ at $68\%$ C.L..

The improvement with respect to \planck\ 2018 legacy release in the large scale polarization data leads to an expected reduction of the $\tau$ uncertainty but it is matched with a slight shift upwards of the central value. This combination leads to a substantial unchanged $\tau$ upper limit leaving to a mostly unchanged constraint on all the minimal $\Lambda$CDM extensions explored. Further investigations are left to future publications. The \srolltwo\ data maps, simulations and, likelihood code are publicly available at \href{http://sroll20.ias.u-psud.fr}{http://sroll20.ias.u-psud.fr}.

\begin{acknowledgements}
We acknowledge the use of CAMB, HEALPix and Healpy software packages. This work is part of the Bware project, partially supported by CNES. It was granted access to the HPC resources of CINES (\url{http://www.cines.fr}) under the allocation 2017-A0030410267 made by GENCI (\url{http://www.genci.fr}). This research used resources of the National Energy Research Scientific Computing Center (NERSC), a U.S. Department of Energy Office of Science User Facility operated under Contract No. DE-AC02-05CH11231. LP is grateful to G. Fabbian, M. Lattanzi and M. Migliaccio for many helpful discussions during the preparation of this work. LP acknowledges the support of the National Centre for Space Studies (CNES) postdoctoral program and Italian Space Agency (ASI) grant 2016-24-H.0 (COSMOS).

\end{acknowledgements}

\bibliographystyle{aat}
\interlinepenalty=10000
\bibliography{bware_tau_paper_final}

\end{document}